\documentclass[varenna]{cimento}
\usepackage{epsfig}
\def\Journal#1#2#3#4{{#1} {\bf #2}, #3 (#4)}

\def\NIM{\em Nucl. Instrum. Methods}

\def\NPB{{\em Nucl. Phys.} B}
\def\PLB{{\em Phys. Lett.}  B}
\def\PRL{\em Phys. Rev. Lett.}
\def\PRD{{\em Phys. Rev.} D}
\def\ZPC{{\em Z. Phys.} C}

\def\be{\begin{equation}}
\def\ee{\end{equation}}
\def\bea{\begin{eqnarray}}
\def\eea{\end{eqnarray}}

\def\etal{{\it et al}}
\begin{document}
\setcounter{page}{0}
\hspace*{\fill}\large HEPSY 97-3\linebreak
\hspace*{\fill}\large Sept. 1997~\linebreak
\vspace*{0.5in}
\centerline{~~~}

\centerline{\Large\bf THE Goals and Techniques of BTeV and LHC-B}
\vspace*{1in}
\centerline{\Large SHELDON STONE}
\vspace*{1in}
\centerline{\it Physics Department, 201 Physics Building, Syracuse Univerisity,\\
Syracuse, NY 13244-1130, USA}
\vspace*{1in}
BTeV and LHC-B are experiments being proposed to study $b$ and
$c$ quark decays in hadron-hadron collisions with the aims to
look for new phenomena beyond the Standard Model and to measure
Standard Model parameters including CKM elements and decay constants. The 
physics goals, required detection techniques and simulations will be discussed.

\vspace*{1in}
\begin{flushleft}
.\dotfill .
\end{flushleft}
\begin{center}
{\large \it Presented at ``Heavy Flavor Physics: A Probe of Nature's Grand Design,"
Varenna, Italy, July 1997} 
\end{center}
\newpage
\normalsize
\title{\bf THE Goals and Techniques of BTeV and LHC-B}
\author{Sheldon Stone}
\institute{Physics Department, 201 Physics Building, Syracuse Univerisity,\\
Syracuse, NY 13244-1130, USA}

\maketitle

\section{Introduction} 
Although most of what is known about $b$ physics
presently has been obtained from $e^+e^-$ colliders operating either at the
$\Upsilon(4S)$ \cite{rankin} or at LEP \cite{roudeau}, interesting information
is now appearing from the hadron collider experiments, CDF and D0, 
which were designed to look
for considerably higher energy phenomena \cite{bedeschi}. The appeal of hadron
colliders arises mainly from the large measured $b$ cross-sections. At the FNAL
collider, 1.8 TeV in the $p\bar{p}$ center-of-mass, the cross-section has been
measured as $\sim$100 $\mu$b \cite{FNALx}, while it is expected to be about five times higher at the
LHC \cite{LHCx}.

At this time there are two proposals for new hadron collider $b$ experiments,
BTeV \cite{BTeV} at the Fermilab collider and LHC-B \cite{LHC-B} at the LHC. Both experiments have
similar physics goals and both want to exploit $b$'s produced in the ``forward"
region as opposed to the current and future centrally foccussed detectors. We
will explore here why such choices were made, and discuss the different
detector designs. We will start with the physics goals.

\section{Physics Goals}
Here we will review what results we want to obtain and what will have already
been accomplished by other experiments before either BTeV or LHC-B starts.

Studies of $b$ and $c$ physics are focused on two main goals. The first is to
look for new phenomena beyond the Standard Model. The second is to measure
Standard Model parameters including CKM elements and decay constants. 

\subsection{The Current Situation}
The Lagrangian for charged current weak decays is
\begin{equation}
L_{cc}=-{g\over\sqrt{2}}J^{\mu}_{cc}W^{\dagger}_{\mu}+h.c., \label{eq:lagrange}
\end{equation}
where
\begin{equation}
J^{\mu}_{cc} =\left(\bar{\nu}_e,~\bar{\nu}_{\mu},~\bar{\nu}_{\tau}\right)
\gamma^\mu \left(\begin{array}{c}e_L\\ \mu_L\\ \tau_L\\\end{array}\right) +
\left(\bar{u}_L,~\bar{c}_{L},~\bar{t}_{L}\right)\gamma^\mu V_{CKM} 
\left(\begin{array}{c}d_L\\  s_L\\ b_L\\ \end{array}\right) 
\end{equation}
and
\begin{equation}
V_{CKM} =\left(\begin{array}{ccc} 
V_{ud} &  V_{us} & V_{ub} \\
V_{cd} &  V_{cs} & V_{cb} \\
V_{td} &  V_{ts} & V_{tb}  \end{array}\right).\end{equation}

Multiplying the mass eigenstates $(d_L, s_L, b_L)$ by the CKM matrix \cite{ckm} leads to the
weak eigenstates. There are nine complex CKM elements. These 18 
numbers can be reduced to four independent quantities by applying unitarity 
constraints and the fact that the phases of the quark wave functions are arbitrary. 
These four remaining numbers are {\bf fundamental constants} of nature that 
need to be determined from experiment, like any other
fundamental constant such as $\alpha$ or $G$. In the Wolfenstein 
approximation the matrix is written in order $\lambda^3$ for the real part
and $\lambda^4$ for the imaginary part as \cite{wolf} 
\begin{equation}
V_{CKM} = \left(\begin{array}{ccc} 
1-\lambda^2/2 &  \lambda & A\lambda^3(\rho-i\eta)(1-\lambda^2/2) \\
-\lambda &  1-\lambda^2/2-i\eta A^2\lambda^4 & A\lambda^2(1+i\eta\lambda^2) \\
A\lambda^3(1-\rho-i\eta) &  -A\lambda^2& 1  
\end{array}\right).
\end{equation}
The constants $\lambda$ and $A$ are determined from charged-current weak 
decays. The measured values are $\lambda = 0.2205\pm 0.0018$ and
A=0.784$\pm$0.043. There are constraints on $\rho$ and $\eta$ from other
measurements that we will discuss. Usually the matrix is viewed only up to
order $\lambda^3$. To explain CP violation in the $K^o$ system the term of
order $\lambda^4$ in $V_{cs}$ is necessary. For the rest of this work, the
higher order terms in $V_{ub}$ and $V_{cb}$ can be ignored.

The unitarity of the CKM matrix\footnote{Unitarity implies that any pair of 
rows or 
columns are orthogonal.} allows us to construct six relationships. The most useful 
turns out to be
\begin{equation}
V_{ud}V_{td}^*+V_{us}V_{ts}^*+V_{ub}V_{tb}^*=0~~.
\end{equation}
To a good approximation
\begin{equation}
V_{ud} \approx V_{tb}^*\approx 1 {\rm ~~ and~~}V_{ts}^*\approx -V_{cb},
\end{equation}
then
\begin{equation}
{V_{ub}\over V_{cb}} + {V_{td}^*\over V_{cb}} - V_{us} = 0~~.
\end{equation}
Since $V_{us}=\lambda$, we can define a triangle with sides
\begin{eqnarray}
1 & & \\ 
\left|{V_{td}\over A\lambda^3 }\right| &=&{1\over \lambda}\sqrt{\left(\rho-
1\right)^2+\eta^2}
={1\over \lambda} \left|{V_{td}\over V_{ts}}\right|\\
\left|{V_{ub}\over A\lambda^3}\right|  &=&{1\over \lambda}\sqrt{\rho^2+\eta^2}
={1\over \lambda} \left|{V_{ub}\over V_{cb}}\right|.
\end{eqnarray}

\begin{figure}[hbtp]
\vspace{-6mm}
\centerline{\epsfig{figure=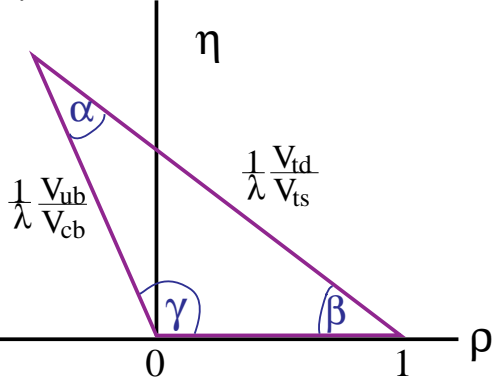,width=3.5in}}
\vspace{-4mm}
\caption{The unitarity triangle shown in the $\rho-\eta$ plane. The
left side is determined by measurements of $b\to u/b\to c$ and the right side can be
determined using mixing measurements in the $B_s$  and $B_d$ systems. The 
angles can be found by making measurements of CP violating asymmetries in
hadronic $B$ decays.
\label{ckm_tri}}
\end{figure}

The CKM triangle is depicted in Fig.~\ref{ckm_tri}. 
We know the length of two sides already:
the base is  defined as unity and the left side is determined by the
measurements of  $|V_{ub}/V_{cb}|$, but the error is still quite
substantial. The right side can be determined using
mixing measurements in the neutral $B$ systems. 
 Fig.~\ref{ckm_tri} also shows the angles
as $\alpha,~\beta$, and $\gamma$. These angles can be determined by measuring
CP violation in the $B$ system. 

\begin{figure}[thb]
\vspace{-.2cm}
\centerline{\epsfig{figure=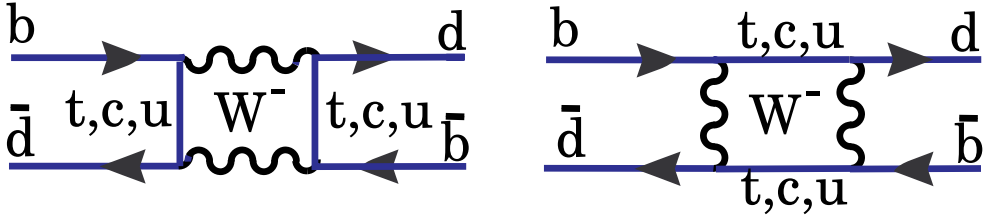,height=1.6in}}
\vspace{-0.6cm}
\caption{\label{bmix}The two diagrams for $B_d$ mixing.} \end{figure}

Neutral $B$ mesons can transform to  their anti-particles before they decay.
The  diagrams for $B_d$ mixing are shown in Fig.~\ref{bmix}. (The diagrams for
$B_s$ mixing are similar with $s$ quarks replacing $d$ quarks.) Although $u$,
$c$ and $t$ quark exchanges are all shown, the $t$ quark plays a dominant role,
mainly due to its mass, since the amplitude of this process is proportional to the
mass of the exchanged fermion. 

 The probability of mixing is given 
by \cite{mixeq}
\begin{equation}
x\equiv \frac{\Delta m}{\Gamma}={G_F^2\over 6\pi^2}B_Bf_B^2m_B\tau_B|V^*_{tb}V_{td}|^2m_t^2
F{\left(m^2_t\over M^2_W\right)}\eta_{QCD},
\end{equation}
where $B_B$ is a parameter related to the probability of the $d$ and $\bar{b}$ 
quarks  forming a hadron and must be estimated theoretically; $F$ is a known
function  which  increases approximately as $m^2_t$, and $\eta_{QCD}$ is a QCD
correction, with value about 0.8. By far the largest uncertainty arises from
the  unknown decay constant, $f_B$. This number gives the coupling between the
$B$ and the $W^-$. It could in principle be determined by finding the decay
rate of $B^+\to \mu^+\nu$ or to $\tau^+\nu$, both of which are very difficult 
to measure.

$B_d$ mixing was first discovered by the
ARGUS  experiment \cite{Argmix}. (There was a previous measurement by UA1
indicating mixing for a mixture of $B_d^o$ and $B_s^o$ \cite{UA1mix}.) At the
time it was quite a surprise, since $m_t$ was thought to be in the 30 GeV
range. Since \begin{equation}  |V^*_{tb}V_{td}|^2\propto
|(1-\rho-i\eta)|^2=(\rho-1)^2+\eta^2, \end{equation} measuring mixing gives a
circle centered at (1,0) in the $\rho - \eta$ plane.

The best recent mixing measurements have been done at LEP, where time-dependent
oscillations have been measured. Averaging over all LEP experiments 
x=0.728$\pm$0.025 \cite{LEPosc}.

\begin{figure}[bht]
\vspace{-.4cm}
\centerline{\epsfig{figure=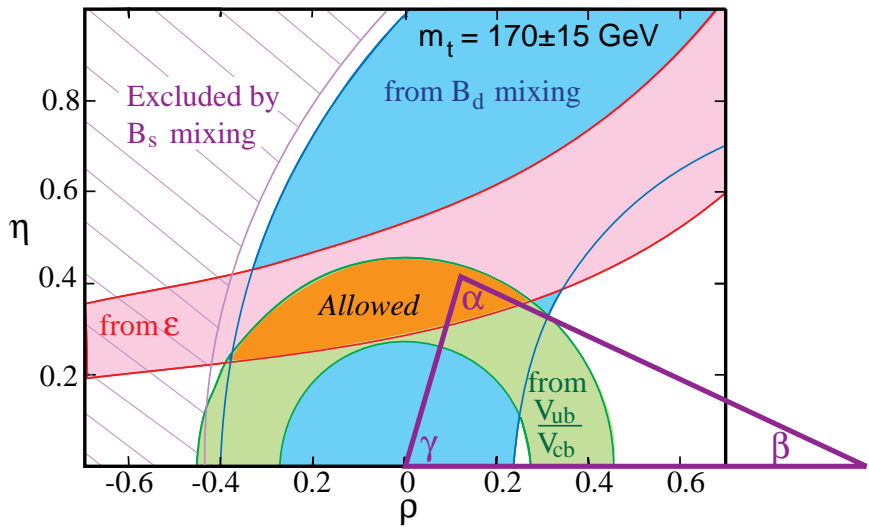,height=3.3in}}
\caption{\label{ckm_fig}The regions in $\rho-\eta$ space (shaded) consistent
with measurements of CP violation in $K_L^o$ decay ($\epsilon$), $V_{ub}/V_{cb}$
in semileptonic $B$ decay, $B_d^o$ mixing, and the excluded region from
limits on $B_s^o$ mixing. The allowed region is defined by the overlap of
the 3 permitted areas, and is where the apex of the  CKM triangle  sits. The
bands represent $\pm 1\sigma$ errors. The large width of the $B_d$ mixing band is
dominated by the uncertainty in the parameter $f_B$. Here the range  is taken as 
$240> f_B > 160$ MeV.}
\end{figure}

Another constraint on $\rho$ and $\eta$ is given by the $K_L^o$ CP
violation  measurement ($\epsilon$) \cite{buras}:
\begin{equation}
\eta\left[(1-\rho)A^2(1.4\pm 0.2)+0.35\right]A^2{B_K \over 0.75}=(0.30\pm 0.06),
\end{equation}
where $B_K$ is parameter that cannot be measured and thus must be calculated.
 I use $0.9>B_k>0.6$ given by an assortment
of theoretical calculations \cite{buras}; this number is one of the largest sources of
uncertainty. Other constraints come from current measurements
on $V_{ub}/V_{cb}$, and $B_d$  mixing \cite{virgin}. The current status of constraints on $\rho$ and $\eta$ is shown in 
Fig.~\ref{ckm_fig}. The width of both of these
bands are also dominated by theoretical errors. Note that the errors used are
$\pm 1\sigma$. This shows that the data are consistent with the standard model
but do not pin down $\rho$ and $\eta$.

It is crucial to check if measurements of the sides and angles are consistent,
i.e., whether or not they actually form a triangle. The standard model is
incomplete. It has many parameters including the four CKM numbers, six quark
masses, gauge boson masses and coupling constants. Perhaps measurements of the
angles and sides of the unitarity triangle will bring us beyond the standard
model and show us how these paramenters are related, or what is missing.

\begin{figure}[htb]
\centerline{\epsfig{figure=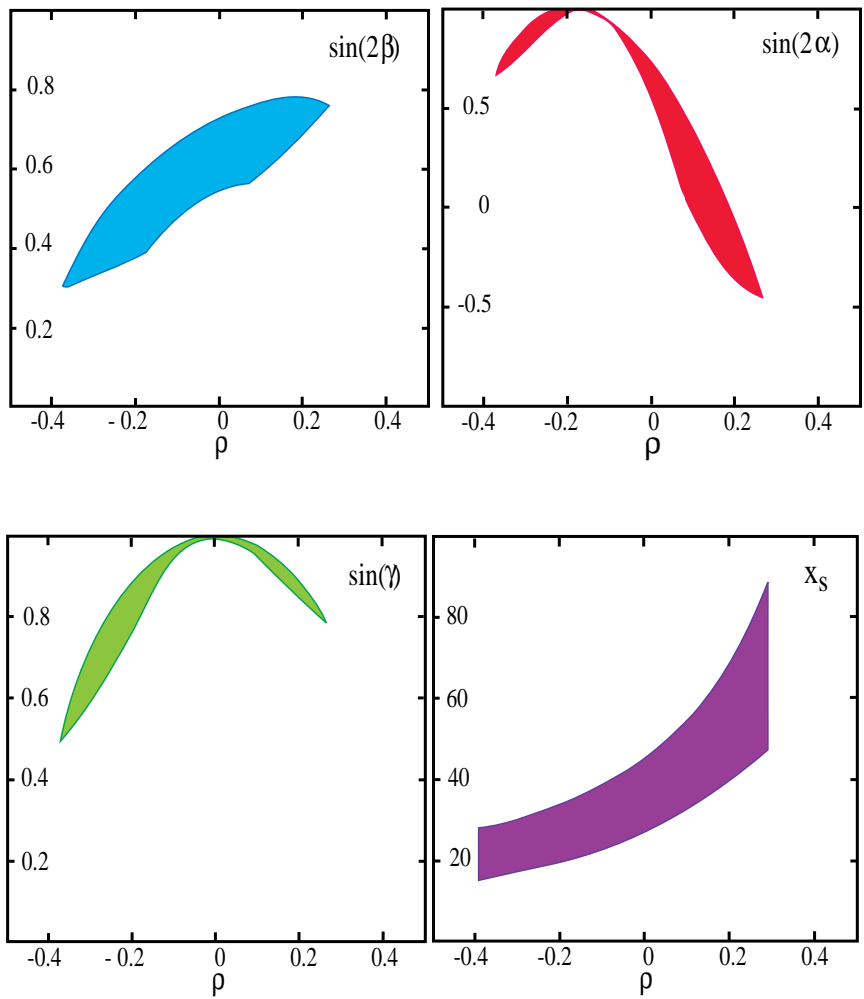,height=5.3in}}
\caption{\label{allowed}The allowed values of three CP violating angles
and the $B_s$ mixing parameter $x_s$ as a function of $\rho$, taken from the
allowed region in Fig.~\ref{ckm_fig}.}
\end{figure}

To get some idea of the size of the effects we need to deal with, I show in
Fig.~\ref{allowed} the expectations for the three CP violating  angles and
$x_s$ plotted versus $\rho$. These plots merely reflect the ``allowed" region
shown in Fig.~\ref{ckm_fig}. It should be emphasized that this is not the
result of a sophisticated analysis,  which is difficult to do because of the
non-Gaussian nature of the theoretical  errors.

Furthermore, new physics can also be observed by measuring branching
ratios which violate standard model predictions. Especially important are
 ``rare decay," processes such as $B\to K\mu^+\mu^-$ or $D\to \pi\mu^+\mu^-$.
 These processes occur only through ``loops," and are a class of what has
 become called ``Penguin" decays. 

\subsection{Formalism of CP Violation in Neutral $B$ Decays}
Consider the operations of Charge Conjugation, C,  and Parity, P:
\begin{eqnarray}
&C|B(\overrightarrow{p})\big>=|\overline{B}(\overrightarrow{p})\big>,~~~~~~~~~
&C|\overline{B}(\overrightarrow{p})\big>=|{B}(\overrightarrow{p})\big> \\
&P|B(\overrightarrow{p})\big>=-|{B}(-\overrightarrow{p})\big>,~~~~~
&P|\overline{B}(\overrightarrow{p})\big>=-|{B}(-\overrightarrow{p})\big> \\
&CP|B(\overrightarrow{p})\big>=-|\overline{B}(-\overrightarrow{p})\big>,~~~
&CP|\overline{B}(\overrightarrow{p})\big>=-|{B}(-\overrightarrow{p})\big> ~~.
\end{eqnarray}
For neutral mesons we can construct the CP eigenstates
\begin{eqnarray}
\big|B^o_1\big>&=&{1\over \sqrt{2}}\left(\big|B^o\big>-
\big|\overline{B}^o\big>\right)~~,\\
\big|B^o_2\big>&=&{1\over 
\sqrt{2}}\left(\big|B^o\big>+\big|\overline{B}^o\big>\right)~~,
\end{eqnarray}
where 
\begin{eqnarray}
CP\big|B^o_1\big>&=&\big|B^o_1\big>~~, \\
CP\big|B^o_2\big>&=&-\big|B^o_2\big>~~.
\end{eqnarray}
Since $B^o$ and $\overline{B}^o$ can mix, the mass eigenstates are a superposition of
$a\big|B^o\big> + b\big|\overline{B}^o\big>$ which obey the Schrodinger equation
\begin{equation}
i{d\over dt}\left(\begin{array}{c}a\\b\end{array}\right)=
H\left(\begin{array}{c}a\\b\end{array}\right)=
\left(M-{i\over 2}\Gamma\right)\left(\begin{array}{c}a\\b\end{array}\right).
\label{eq:schrod}
\end{equation}
If CP is not conserved then the eigenvectors, the mass eigenstates $\big|B_L\big>  $ 
and  $\big|B_H\big>$, are not the CP eigenstates but are 
\begin{equation}
\big|B_L\big> = p\big|B^o\big>+q\big|\overline{B}^o\big>,~~\big|B_H\big> = 
p\big|B^o\big>-q\big|\overline{B}^o\big>,
\end{equation}
where
\begin{equation}
p={1\over \sqrt{2}}{{1+\epsilon_B}\over {\sqrt{1+|\epsilon_B|^2}}},~~
q={1\over \sqrt{2}}{{1-\epsilon_B}\over {\sqrt{1+|\epsilon_B|^2}}}.
\end{equation}
CP is violated if $\epsilon_B\neq 0$, which occurs if $|q/p|\neq 1$.

The time dependence of the mass eigenstates is 
\begin{eqnarray}
\big|B_L(t)\big> &= &e^{-\Gamma_Lt/2}e^{im_Lt/2} \big|B_L(0)\big> \\
 \big|B_H(t)\big> &= &e^{-\Gamma_Ht/2}e^{im_Ht/2} \big|B_H(0)\big>,
\end{eqnarray}
leading to the time evolution of the flavor eigenstates as
\begin{eqnarray}
\big|B^o(t)\big>&=&e^{-\left(im+{\Gamma\over 2}\right)t}
\left(\cos{\Delta mt\over 2}\big|B^o(0)\big>+i{q\over p}\sin{\Delta mt\over 
2}\big|\overline{B}^o(0)\big>\right) \\
\big|\overline{B}^o(t)\big>&=&e^{-\left(im+{\Gamma\over 2}\right)t}
\left(i{p\over q}\sin{\Delta mt\over 2}\big|B^o(0)\big>+
\cos{\Delta mt\over 2}\big|\overline{B}^o(0)\big>\right),
\end{eqnarray}
where $m=(m_L+m_H)/2$, $\Delta m=m_H-m_L$ and 
$\Gamma=\Gamma_L\approx \Gamma_H$, and $t$ is the decay time in the $B^o$
rest frame, the so called proper time. Note that the probability of a 
$B^o$ decay as a function of $t$ is given by $\big<B^o(t)\big|B^o(t)\big>^*$, and is 
a pure exponential, $e^{-\Gamma t/2}$, in the absence of CP violation.

\subsubsection{CP violation for $B$ via interference of mixing and decays}

\begin{figure}[htb]
\centerline{\epsfig{figure=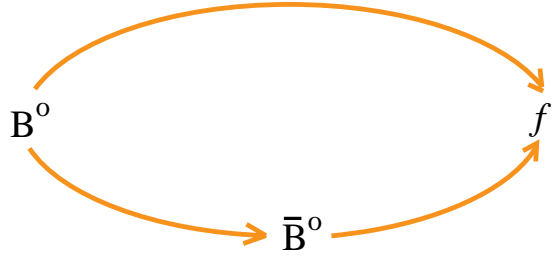,height=1.5in}}
\caption{\label{eigen_CP}Two interfering ways for a $B^o$ to decay into a final
state $f$.} 
\end{figure}

Here we choose a final state $f$ which is accessible to both $B^o$ and
$\overline{B}^o$  decays \cite{BigiSanda}. The second amplitude necessary for
interference is provided by mixing.  Fig.~\ref{eigen_CP} shows the decay into
$f$ either directly or indirectly via  mixing. 
It is  necessary only that $f$ be
accessible directly from either state; however if $f$ is a CP  eigenstate the
situation is far simpler. For  CP eigenstates \begin{equation}
CP\big|f_{CP}\big>=\pm\big|f_{CP}\big>. \end{equation}

It is useful to define the amplitudes
\begin{equation}
A=\big<f_{CP}\big|{\cal H}\big|B^o\big>,~~
\overline{A}=\big<f_{CP}\big|{\cal H}\big|\overline{B}^o\big>.
\end{equation}
If $\left|{\overline{A}\over A}\right|\neq 1$, then we have ``direct" CP violation in the 
decay 
amplitude, which we will discuss in detail later. Here CP can be violated by having
\begin{equation}
\lambda = {q\over p}\cdot {\overline{A}\over A}\neq 1,
\end{equation}
which requires only that $\lambda$  acquire a non-zero phase, i.e. $|\lambda|$ 
could be 
unity and CP violation can occur. 

A comment on neutral $B$ production at $e^+e^-$ colliders is in order. At the 
$\Upsilon 
(4S)$ resonance there is coherent production of $B^o\bar{B}^o$ pairs. This puts the 
$B$'s in a $C=-1$ state. In hadron colliders, or at $e^+e^-$ machines operating 
at the $Z^o$, the $B$'s are produced incoherently \cite{BigiSanda}. For the rest
of this article I will 
assume incoherent production except where explicitly noted.

The asymmetry, in this case, is defined as
\begin{equation}
a_{f_{CP}}={{\Gamma\left(B^o(t)\to f_{CP}\right)- 
\Gamma\left(\overline{B}^o(t)\to 
f_{CP}\right)}\over
{\Gamma\left(B^o(t)\to f_{CP}\right)+ \Gamma\left(\overline{B}^o(t)\to 
f_{CP}\right)}},
\end{equation}
which for $|q/p|=1$ gives
\begin{equation}
a_{f_{CP}}={{\left(1-|\lambda|^2\right)\cos\left(\Delta mt\right)-2{\rm Im}\lambda 
\sin(\Delta mt)}\over {1+|\lambda|^2}}.
\end{equation}
For the cases where there is only one decay amplitude $A$, $|\lambda |$ equals 1, 
and we have
\begin{equation}
a_{f_{CP}}=-{\rm Im}\lambda \sin(\Delta mt).
\end{equation}
Only the amplitude, ${\rm -Im}\lambda$ contains information about the level of CP 
violation, 
the sine term is determined only by $B_d$ mixing. In fact, the time integrated 
asymmetry is given by
\begin{equation}
a_{f_{CP}}=-{x \over {1+x^2}}{\rm Im}\lambda = -0.48 {\rm Im}\lambda ~~. \label{eq:aint}
\end{equation}
This is quite lucky as the maximum size of the coefficient for any $x$ is $-0.5$.

Let us now find out how ${\rm Im}\lambda$ relates to the CKM parameters. Recall 
$\lambda={q\over p}\cdot {\overline{A}\over A}$. The first term is the part that comes 
from mixing:
\begin{equation}
{q\over p}={{\left(V_{tb}^*V_{td}\right)^2}\over {\left|V_{tb}V_{td}\right|^2}}
={{\left(1-\rho-i\eta\right)^2}\over {\left(1-\rho+i\eta\right)\left(1-\rho-
i\eta\right)}}
=e^{-2i\beta}{\rm~~and}
\end{equation}
\begin{equation}
{\rm Im}{q\over p}= -{{2(1-\rho)\eta}\over {\left(1-\rho\right)^2+\eta^2}}=\sin(2\beta).
\end{equation}

\begin{figure}[htb]
\vspace{-.4cm}
\centerline{\epsfig{figure=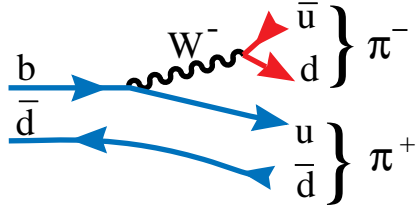,height=1.7in}}
\vspace{-.5cm}
\caption{\label{pippim}Decay diagram at the tree level for $B^o\to\pi^+\pi^-$.}
\end{figure}

To evaluate the decay part we need to consider specific final states. For example, 
consider 
$f\equiv\pi^+\pi^-$. The simple spectator decay diagram is shown in 
Fig.~\ref{pippim}.
For the moment we will assume that this is the only diagram which contributes. 
Later I will 
show why this is not true. For this $b\to u\bar{u}d$ process we have
\begin{equation}
{\overline{A}\over A}={{\left(V_{ud}^*V_{ub}\right)^2}\over 
{\left|V_{ud}V_{ub}\right|^2}}={{(\rho-i\eta)^2}\over
 {(\rho-i\eta)(\rho+i\eta)}}=e^{-2i\gamma},
\end{equation}
and 
\begin{equation}
{\rm Im}(\lambda)={\rm Im}(e^{-2i\beta}e^{-2i\gamma})=
{\rm Im}(e^{2i\alpha})=-\sin(2\alpha).
\end{equation}

\begin{figure}[htb]
\vspace{-.4cm}
\centerline{\epsfig{figure=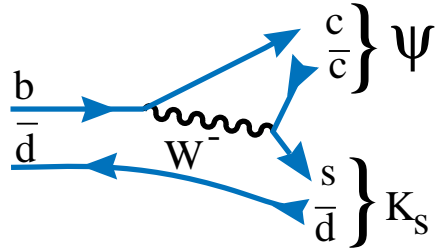,height=2.0in}}
\vspace{-.15cm}
\caption{\label{psi_ks}Decay diagram at the tree level for $B^o\to\psi K_S$.}
\end{figure}

For our next example let's consider the final state $\psi K_S$. The decay diagram is 
shown in Fig.~\ref{psi_ks}. In this case we do not get a phase from the decay part because
\begin{equation}
{\overline{A}\over A} = {{\left(V_{cb}V_{cs}^*\right)^2}\over 
{\left|V_{cb}V_{cs}\right|^2}}
\end{equation}
is real to order $1/\lambda^4$.
In this case the final state is a state of negative $CP$, i.e. 
$CP\big|\psi K_S\big>=-\big|\psi K_S\big>$. This introduces an additional
minus sign in the result for ${\rm Im}\lambda$.
Before finishing discussion of this final state we need to consider in more detail 
the presence of the $K_S$ in the final state. Since neutral kaons can mix, we pick 
up another mixing phase (similar diagrams as for $B^o$, see Fig.~\ref{bmix}). 
This term creates a phase given by
\begin{equation}
\left({q\over p}\right)_K={{\left(V_{cd}^*V_{cs}\right)^2}\over 
{\left|V_{cd}V_{cs}\right|^2}},
\end{equation}
which is real to order $\lambda^4$. It necessary to include this term, however,
since there are other  formulations of the CKM matrix than Wolfenstein, which
have the phase in a  different location. It is important that the physics
predictions not depend on the  CKM convention.\footnote{Here we don't include
CP violation in the neutral kaon  since it is much smaller than what is
expected in the $B$ decay. The term of order $\lambda^4$ in $V_{cs}$ is
necessary to explain $K^o$ CP violation.}

In summary, for the case of $f=\psi K_S$, ${\rm Im}\lambda=-\sin(2\beta)$.
\subsubsection{Comment on Penguin Amplitude}

\begin{figure}[htb]
\vspace{-.8cm}
\centerline{\epsfig{figure=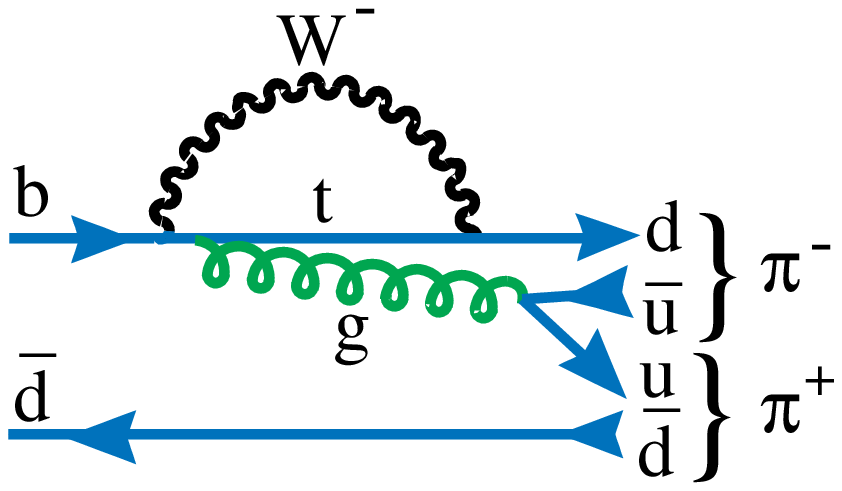,height=1.7in}}
\vspace{-.2cm}
\caption{\label{pipi_penguin}Penguin diagram for $B^o\to\pi^+\pi^-$.}
\end{figure}

In principle all processes can have penguin components. One such diagram is
shown  in Fig.~\ref{pipi_penguin}. The $\pi^+\pi^-$ final state is expected to
have a rather large penguin  amplitude $\sim$10\% of the tree amplitude. Then
$|\lambda |\neq 1$ and  $a_{\pi\pi}(t)$ develops a $\cos(\Delta mt)$ term. It
turns out that $\sin(2\alpha)$ can be extracted using isospin considerations
and  measurements of the branching ratios for $B^+\to \pi^+\pi^o$ and $B^o\to 
\pi^o\pi^o$, or other methods \cite{extract_alpha}.

In the $\psi K_S$ case, the penguin amplitude is expected to be small since a 
$c\bar{c}$ pair must be ``popped" from the vacuum. Even if the penguin decay 
amplitude were of significant size, the decay phase is the same as the tree level 
process, namely zero.

\subsection{Charm Physics Goals}

According to the standard model, charm mixing and CP violating effects should
be ``small." Thus charm provides an excellent place for non-standard model
effects to a appear. Specific goals are listed below.

$\bullet$ Search for mixing in $D^o$ decay, by looking for both the rate of 
wrong sign decay, $r_D$ and the width difference between positive CP and
negative CP eigenstate decays, $\Delta\Gamma$. The current upper limit on $r_D$
is $3.7\times 10^{-3}$, while the standard model expectation is
$r_D<10^{-7}$ \cite{dmix}.

$\bullet$ Search for CP violation in $D^o$. Here we have the advantage over $b$
decays that there is a large $D^{*+}$ signal which tags the inital flavor of
the $D^o$ through the decay $D^{*+}\to \pi^+ D^o$. Similarly $D^{*-}$ decays
tag the flavor of inital $\overline{D}^o.$ The current experimental upper 
limits on CP violating asymmetries are on the order of 10\%, while the standard
model prediction is about 0.1\% \cite{dcp}.

$\bullet$ We also want to measure $\Delta\Gamma$, the lifetime difference
between CP+ and CP- eigenstates. To do this we can measure the lifetime in the
decay mode $D^o\to K^+K^-$, which is pure CP+ and compare it with the lifetime
measured using $D^o\to K^-\pi^+$, which is an equal mixture of CP states.

$\bullet$ Search for rare decays of charm, which if found would signal new
physics.

\subsection{$b$ Physics Goals}
We expect that the phase of $B_d$ mixing, called $\sin(2\beta)$, will have been
measured prior to the advent of BTeV or LHC-B by at least one of the 
experiments: Babar, Belle, CDF or HERAB \cite{Schleins}. There is also a possibility of that a
CP violating rate in flavor specific decays such as $B\to K^-\pi^+$ will have
been seen. We do not however, expect that measurements will exist of our main
physics goals, which are listed below:

$\bullet$ Measurement of the CP violating asymmetry in $B^o\to \pi^+\pi^-$.
Penguin pollution may cause a problem for the extraction of the angle $\alpha$.
There have been several theoretical suggestions of how to do this using
additional branching ratios measurements \cite{extract_alpha}.

\begin{figure}[htb]
\centerline{\epsfig{figure=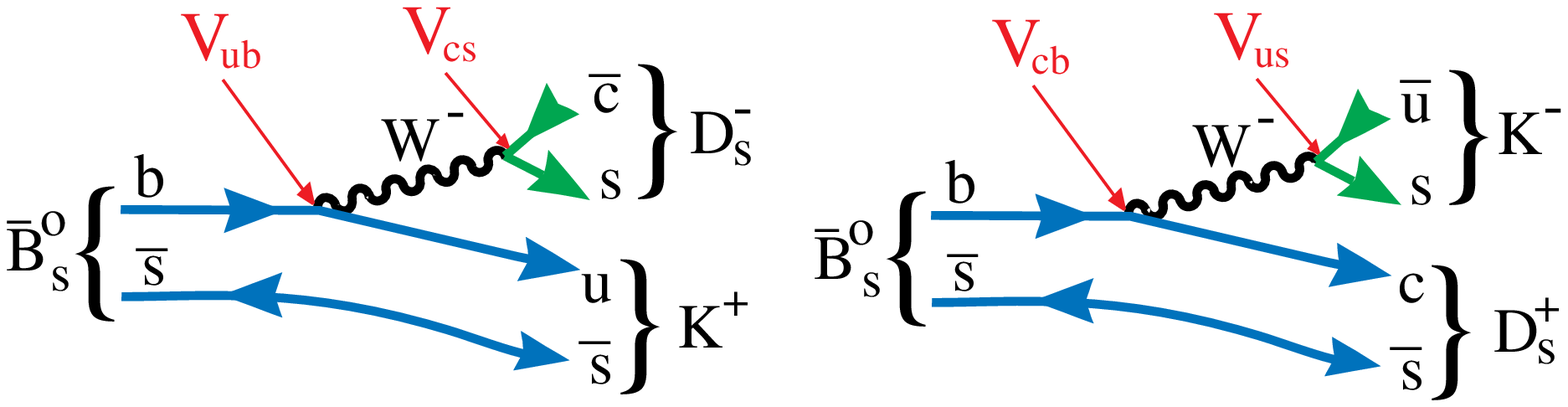,height=1.8in}}
\vspace{-.15cm}
\caption{\label{DsK} Two diagrams for $\overline{B}_s^o \to D_s^{\pm}K^{\mp}$.}
\end{figure}

$\bullet$ Measurment of  mixing and CP violation in $B_s$ decays. The mass
difference between ``short" and ``long" eigenstates can be determined measuring
the flavor tagged time dependence of one or more of the final states $\psi
K^{*o}$, $D_s^+\pi^-$ or $D_s^+\pi^+\pi^-\pi^-$.	It is also possible
that there is sizable $\Delta\Gamma$ in the $B_s$ system. Some estimates give 
$\Delta\Gamma \sim 0.1\times\Gamma$ \cite{DeltaGam}. This can be determined by
measuring the lifetimes separately for CP+ and CP- eigenstates. For example,
$D_s^+D_s^-$ is a CP+ eigenstate, while $\psi K_s$ is a CP- eigenstate. These
final states are useful to look for a large non-standard model time dependent
CP violating asymmetry, since in the standard model only small asymmetry is
expected. Exquisit time resolution is required, since $B_s$ oscillation are
expected to be quite rapid. Very importantly, time dependent measurements of
the decays to the final states $D_s^{\pm}K^{\mp}$ can provide a direct
measurement of the angle $\gamma$ \cite{CPgam}.  Fig.~\ref{DsK} shows the two
direct decay processes for $\overline{B}^o_s$. These interfere with the mixing
induced processes.

$\bullet$ Measurement of rare decay processes such as $B^o\to K^+\pi^-$, $B\to
K\ell^+\ell^-$, etc...

$\bullet$ Precision measurement of $\sin (2\beta)$ using $\psi K_s$.

$\bullet$ Precision measurements of $V_{ub}/V_{cb}$ and $V_{cb}$ using both
meson and baryon decays such as $\Lambda_b\to p\ell\nu.$

\section{Characteristics of Hadronic $b$ Production}

It is often customary to characterize heavy quark production in hadron
collisions with the two variables $p_t$ and $\eta$. The later variable was
first invented by those who studied high energy cosmic rays and is assigned the
value 
\begin{equation}
\eta = -ln\left(\tan\left({\theta/2}\right)\right),
\end{equation}
where $\theta$ is the angle of the particle with respect to the beam direction.

\begin{figure}[htb]
\vspace{-.9cm}
\epsfig{figure=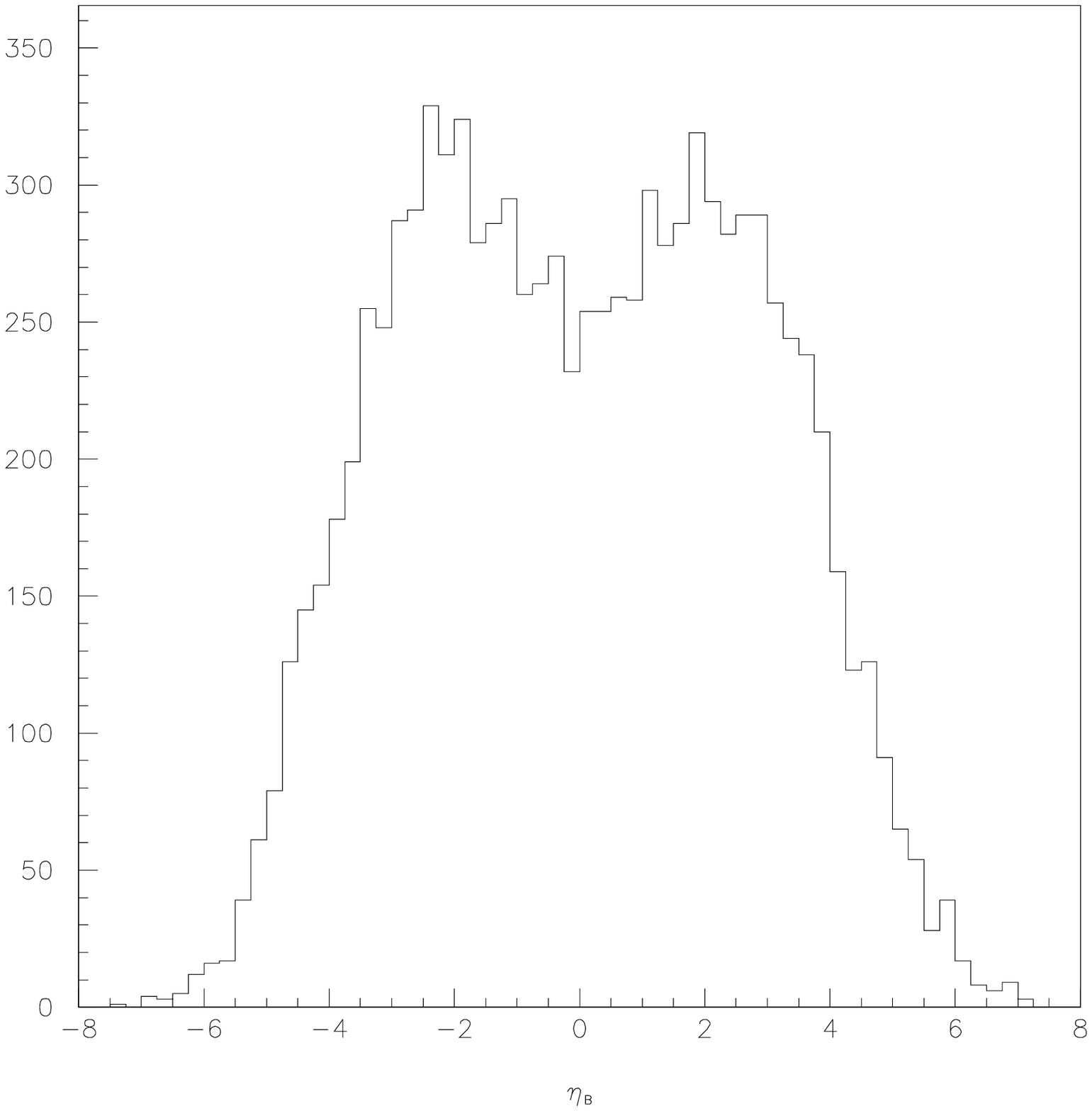,width=2.5in}\
\epsfig{figure=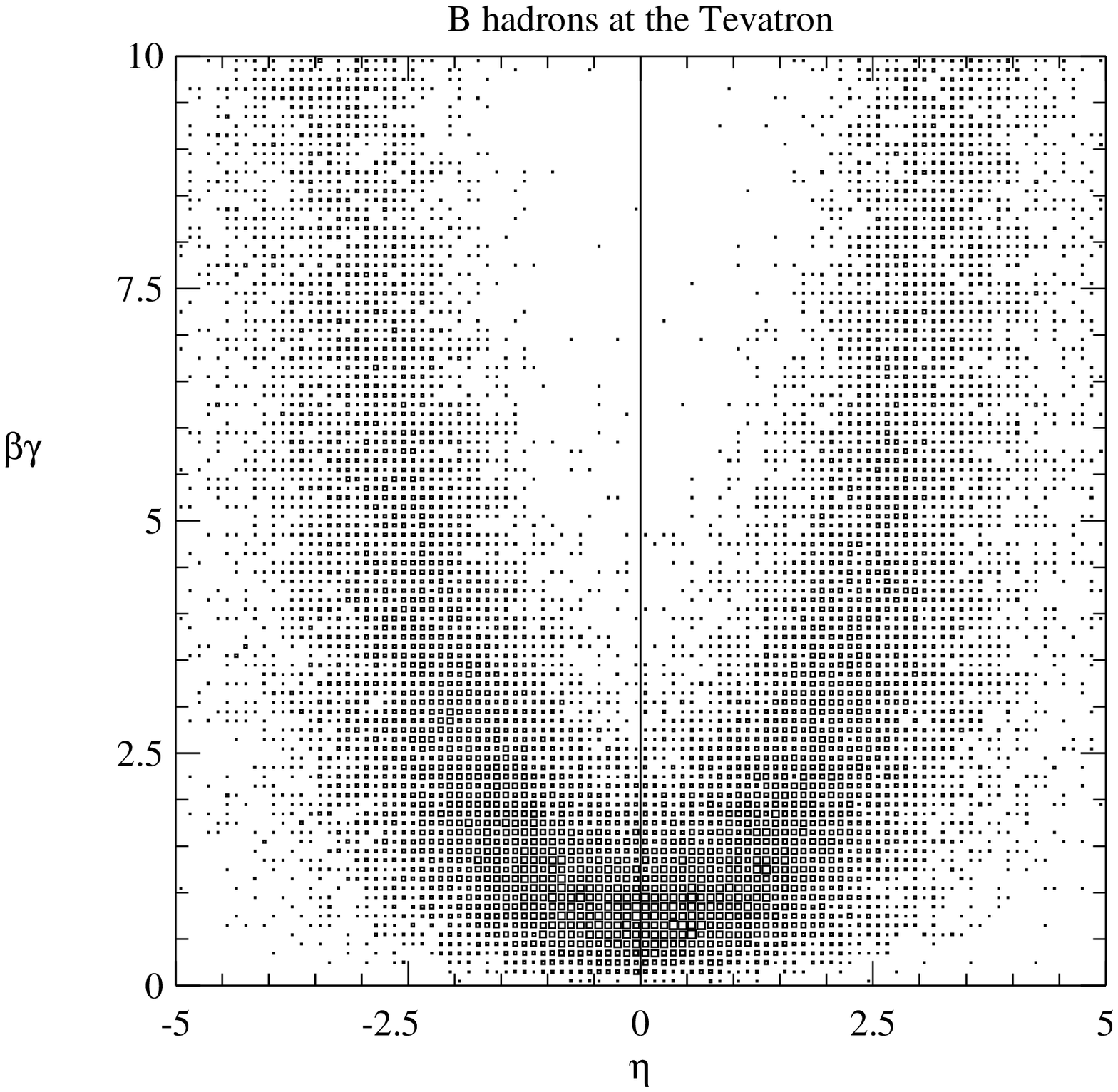,width=2.6in}
\caption{\label{n_vs_eta}  The $B$ yield versus $\eta$ (left). 
$\beta\gamma$ of the  $B$  versus $\eta$ (right). Both plots are for the
Tevatron.}
\end{figure}

According to QCD based calculations of $b$ quark production, the $b$'s are
produced ``uniformly" in $\eta$ and have a truncated transverse momentum,
$p_t$, spectrum, characterized by a mean value approximately equal to the $B$
mass \cite{artuso}. The distribution in $\eta$ is shown in Fig.~\ref{n_vs_eta}.

\begin{figure}[htb]
\vspace{-4.2cm}
\centerline{\epsfig{figure=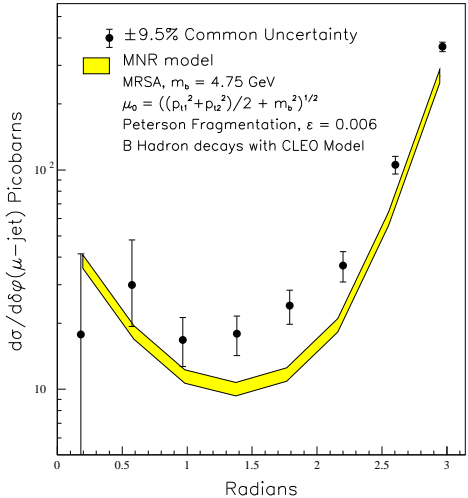,height=8.in}}
\vspace{-9.cm}
\caption{\label{cdf_dphi}The differential $\delta\phi$ cross-sections for
$p^{\mu}_T> 9 $ GeV/c, $\left|\eta^{\mu}\right|<$0.6, E$^{\bar{b}}_T>$10 GeV,
$\left|\eta^{\bar{b}}\right|<1.5$ compared with theoretical predictions. The
data points have a common systematic uncertainty of $\pm$9.5\%. The uncertainty
in the theory curve arises from the error on the muonic branching ratio and
the uncertainty in the fragmentation model.}
\end{figure}

There is a strong correlation between the $B$ momentum and $\eta$. Shown also in
Fig.~\ref{n_vs_eta} is the $\beta\gamma$ of the $B$ hadron versus $\eta$.
It can clearly be seen that near $\eta$ of zero, $\beta\gamma\approx 1$, while
at larger values of $|\eta |$, $\beta\gamma$ can easily reach values of ~6.
This is important because the observed decay length varies with $\beta\gamma$
and furthermore the absolute momenta of the decay products are larger allowing
for a supression of the multiple scattering error.

Since the detector design is somewhat dependent on the Monte Carlo generated
$b$ production distributions, it is important to check that the correlations
between the $b$ and the $\overline{b}$ are adequately reproduced. 
 In Fig.~\ref{cdf_dphi} I show the 
azimuthal opening angle distribution between a muon from a $b$ quark decay and the
$\bar{b}$ jet as measured by CDF~\cite{cdf_prod} and 
compare with the MNR predictions \cite{MNR}.
 
The model does a good job in
representing the shape which shows a strong back-to-back correlation. The
normalization is about a factor of two higher in the data than the theory,
which is generally
true of CDF $b$ cross-section measurements \cite{cdf_bx}. 
In hadron colliders all $B$ species are produced at the same time.

\begin{figure}[htb]
\centerline{\epsfig{figure=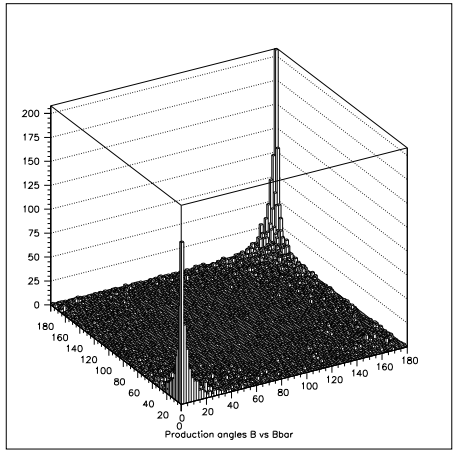,height=3.in}}
\vspace{-.6cm}
\caption{\label{bbar}The production angle (in degrees) for the hadron
containing a $b$ quark plotted versus the production angle for a hadron
containing $\bar{b}$ quark. (For the Tevatron.)}
\end{figure}

The ``flat" $\eta$ distribution hides an important correlation of
$b\bar{b}$ production at hadronic colliders. In Fig.~\ref{bbar} the production
angles of the hadron containing the $b$ quark is plotted versus the production
angle of the hadron containing the $\bar{b}$ quark according to the Pythia
generator (for the Tevatron). There is a very strong
correlation in the forward (and backward) direction: when the $B$ is forward
the $\overline{B}$ is also forward. This correlation is not present in the
central region (near zero degrees). By instrumenting a relative small region of
angular phase space, a large number of $b\bar{b}$ pairs can be detected.
Furthermore the $B$'s populating the forward and backward regions have large
values of $\beta\gamma$. 

Charm production is similar to $b$ production but much larger. Current
theoretical esitmates are that charm is 1-2\% of the total $p\bar{p}$
cross-section. Table ~\ref{tab:b_c} gives the relevant parameters for the
Tevatron and LHC.
BTeV expecst to start serious data taking in Fermilab Run II with a luminosity of
about $5\times 10^{31}$cm$^{-2}$s$^{-1}$; the ultimate luminosity goal, to be
obtained in Run III is $2\times 10^{32}$cm$^{-2}$s$^{-1}$. 
\begin{table}[t]
\caption{Properties of the Tevatron and LHC as a $b$ and $c$ sources.\label{tab:b_c}}
\vspace{0.4cm}
\begin{center}
\begin{narrowtabular}{1.5cm}{lcc}   \hline
Property  & Fermilab (Run II) & LHC-B\\\hline
Luminosity & $0.5\times 10^{32}$cm$^{-2}$s$^{-1}$ $[\dagger]$ & 
$1.5\times 10^{32}$cm$^{-2}$s$^{-1}$\\
$b$ cross-section & 100 $\mu$b  & 500 $\mu$b\\ 
\# of $b$'s per 10$^7$ sec  & $10^{11}$ &$5\times 10^{11}$ \\
$b$ fraction & 0.2\%  & 0.5\%\\
$c$ cross-section & $>500~\mu$b &$>4000~\mu$b\\
Bunch spacing & 132 ns & 25 ns \\
Luminous region length & $\sigma_z$ = 30 cm & 5.3 cm\\
Luminous region width & $\sigma_x$=$\sigma_y$ $\approx$ 50 $\mu$m & 
 $\sigma_x$=$\sigma_y$ = 70 $\mu$m\\
Interactions/crossing & $<0.5>$ & $<0.4>$
 \\ \hline
\end{narrowtabular}
\end{center}
$\dagger$ The ultimate luminosity for BTeV is projected to be $2\times
10^{32}$cm$^{-2}$s$^{-1}$.
\end{table}

\section{Designs of LHC-B and BTeV}

\subsection{Introduction}

\begin{figure}[htb]
\vspace{-2.5cm}
\centerline{\epsfig{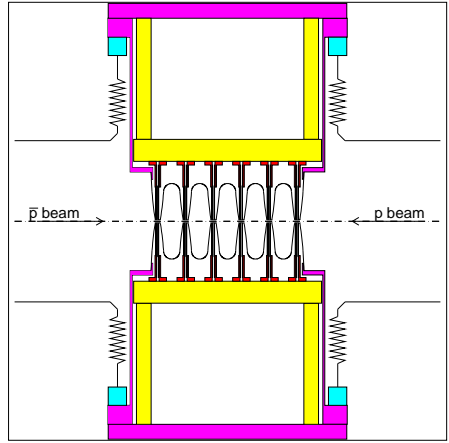}}
\vspace{-.05cm}
\caption{\label{p238}Side view of the P238 silicon detector
assembly and Roman pots. The 6 silicon planes are the vertical lines just above
and below the beam line. The bellows (zig-zag lines) allow movements in the
vertical direction of the pots, which are the thin vertical lines close to the
bellows (they have 2 mm wall thickness). The edges of the 200 $\mu$m-thick
aluminum RF shields closest to the beam (shown as the thin curved lines near
the silicon detectors) normally ran at a distance of 1.5 mm from the
circulating beams. The black horizontal pieces at top and bottom are the vacuum
bulkheads bolted to the Roman pots. }
\end{figure}

Both dedicated $B$ detectors have decided to use the ``forward" direction.
There are several reasons for this. First of all, it is possible to get a
signficant fraction of events with both the $b$ and $\overline{b}$ in the
detector. Secondly, the geometry allows space for charged particle
identification. Thirdly, it is possible to place a microvertex detector inside
the main beam pipe and retract it during machine fills, necessary to minimize
radiation damage.  A
test of this concept was done at CERN by experiment P238 \cite{p238p}. A sketch
of their silicon detector arrangement is shown in
Fig.~\ref{p238}.

Since in hadron-hadron interactions there are many more hadronic interactions
than interesting events, it is necessary to select which events to keep for
further processing. In normal parlance this is called ``triggering the
experiment." The implementation of the trigger is crucial to the success of any
hadronic $b$ experiment. BTeV is designed around the ability to trigger on the
detached vertex created by the decay of hadron containing a $b$ or $c$ quark.
BTeV and LHC-B also have the ability to trigger on events with opposite sign
$e^+e^-$ or $\mu^+\mu^-$ pairs. Because of the specific design of its
microvertex detector, LHC-B cannot do a detached vertex trigger in the first
level lest they be swamped with too high a trigger rate \cite{LHCB_trig}. They
require that they have either electrons, muons or hadrons at moderate $p_t$
before they look for the consequences of a detached vertex. These requirement
causes a loss of efficiency which depends on the charged multiplicity of the
final state.

Both experiments have excellent charged hadron identification, and lepton
identification as well as excellent momentum resolution. Neither experiment has
investigated in detail the efficacy of using photons or neutral pions. I will
assume, conservatively, in this paper that they cannot be used.

 \subsection{The BTeV Design}
\begin{figure}[htbp]
\centerline{\epsfig{figure=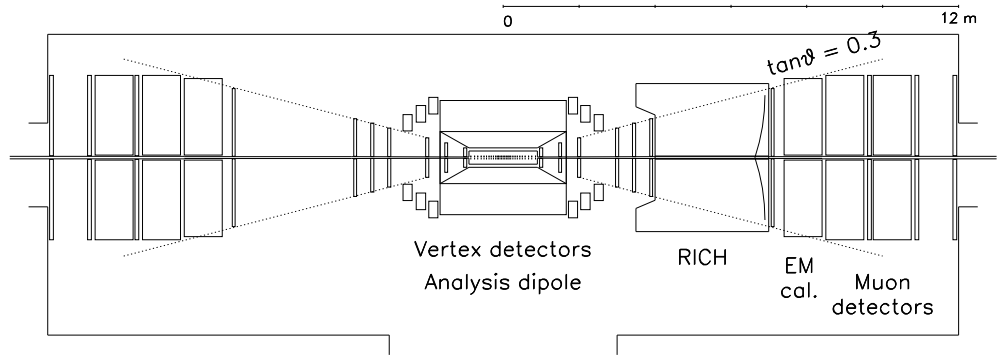,height=2.3in}}
\caption{\label{btev_det_doc}Sketch of the
BTeV spectrometer.}
\end{figure}

 A sketch of the BTeV apparatus is shown in Fig.~\ref{btev_det_doc}. The plan view
shows the two-arm spectrometer fitting in the expanded C0 interaction region
at Fermilab. The magnet, called SM3, exists at Fermilab. The other important parts of the
experiment include the vertex detector, the RICH detectors, the EM calorimeters
and the muon system. The solid angle subtended is approximately $\pm$300 mr in both plan and
elevation views. 

The C0 interaction region at Fermilab will be enlarged and a counting room
constructed 
starting around Nov. 15, 1997. It will be completed before the main injector turns on.

\subsubsection{The BTeV pixel vertex detector}
\begin{figure}[htbp]
\centerline{\epsfig{figure=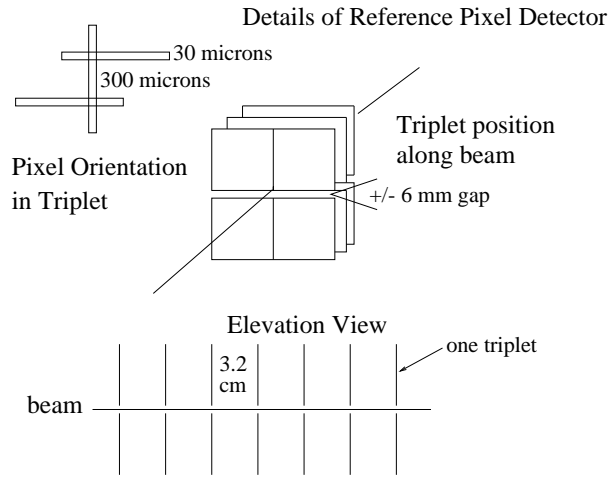,height=3.1in,angle=90}}
\caption{\label{vertex}Sketch of the pixel vertex detector for BTeV.}
\end{figure}

The vertex detector is a multiplane pixel device which sits
inside the beam pipe. A sketch is shown in Fig.~\ref{vertex}. This detector is
the most versatile part of the experiment. It is used for precise measurement
of the vertex position of the primary $p\overline{p}$ interaction and heavy
quark decay vertices. It is the key element of the detached vertex trigger and
it is an integral part of the charged particle momentum measuring system. 

Data is ``pipelined" out of the detector and kept in temporary storage until a
decision can be made on wether or not there is a 
 detached $b$ or $c$
vertex. The vertex detector is put in the
magnetic field in order to insure that the tracks considered for vertex based
triggers do not have large mulitple scattering because they are low momentum.

Simulations have been carried out taking as a baseline detector triplets of
pixels with dimension 30 $\mu$m x 300 $\mu$m. The function of the triplet is to
provide a track vector in the bend plane, along with a three dimensional
spacial coordinate. This useful in making the trigger algorithm fast. These
pixel dimensions are the minimum BTeV is considering using. Studies are
underway to see how much bigger the pixels can be without degrading the
performance. Also, the minimum distance of the detector to the beam line is set
at 6 mm, though the beam size is small enough that a smaller distance could be
considered.

In addition to the pixel detector, the charged particle tracking system
contains several planes of wire or straw chambers. These are particularly
necessary to provide excellent momentum resolution and to reconstruct $K_s^o$
decays into $\pi^+\pi^-$

\subsubsection{Particle Identification}

Excellent charged hadron identification is a critical component of a 
heavy quark experiment. Even for a spectrometer with the excellent mass 
resolution of BTeV, there are kinematic regions where signals from 
one final state 
will overlap those of another final state. For example, 
$B_{d}^{o}\rightarrow \pi^{+}\pi^{-}$, $B_{d}^{o}\rightarrow K^{+}\pi^{-}$,
and $B_{s}^{o}\rightarrow K^{+}K^{-}$ all overlap to some degree. 
These ambiguities can be eliminated almost 
entirely by an effective particle identifier. In addition, many physics
investigations involving neutral $B$-mesons require `tagging' of the flavor 
of the signal particle by examining the properties of the `away-side' 
particle. Kaon tagging is a very effective means of 
doing this.

\begin{figure}[hbt]
\centerline{ \epsfxsize=2.6in \epsffile{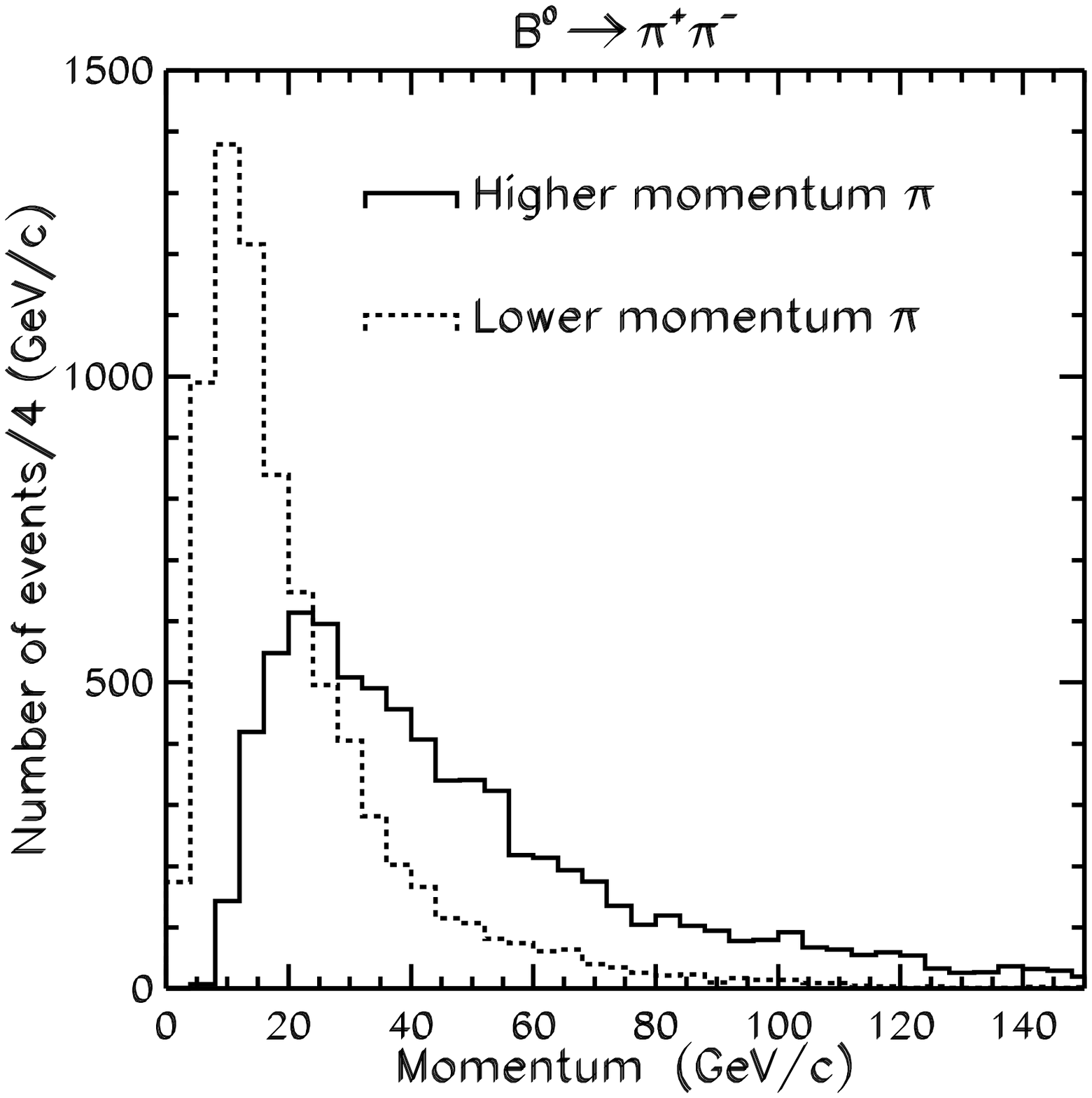}
             \epsfxsize=2.6in \epsffile{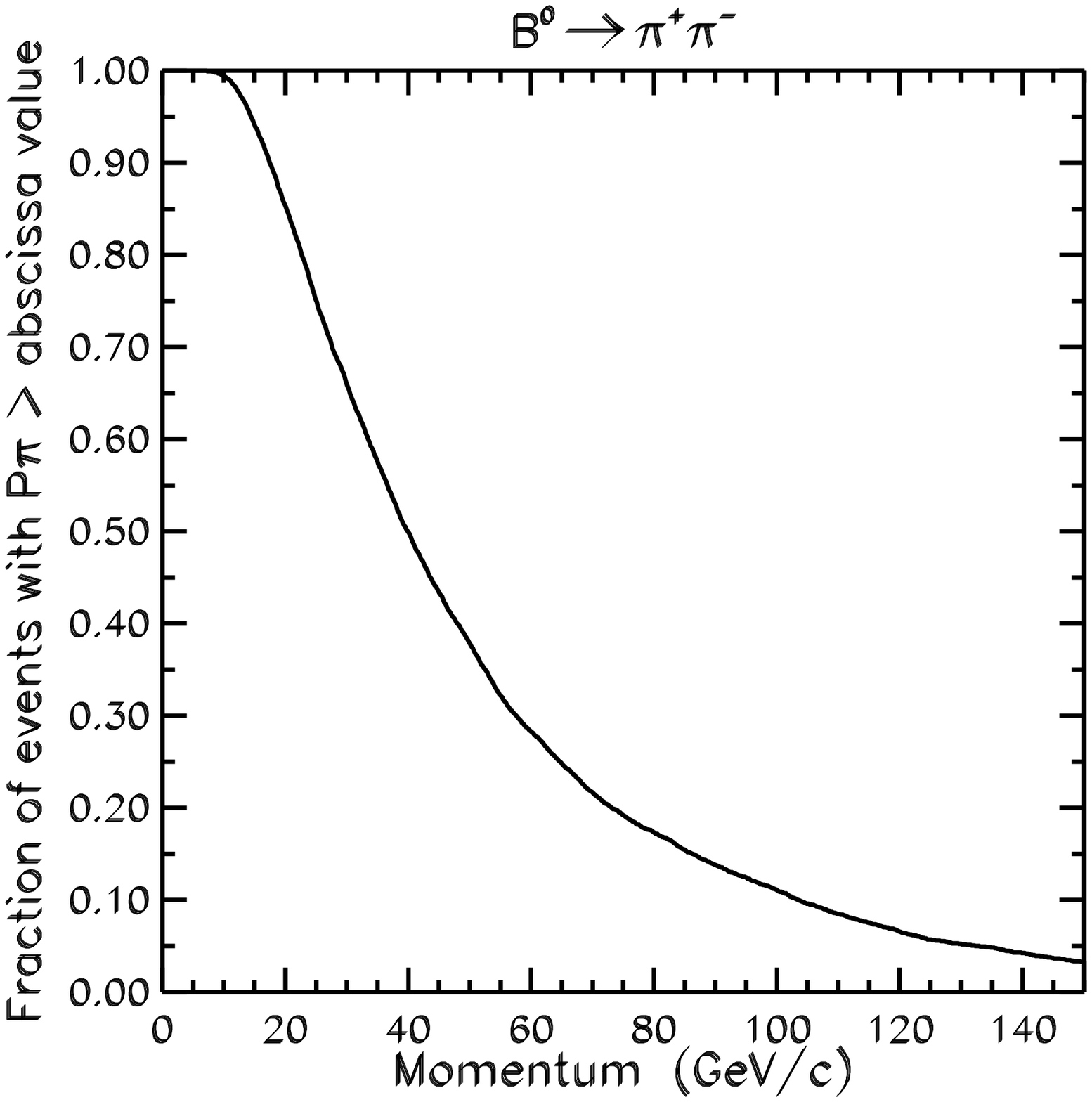}} 
\caption{The momentum distribution of pions in $B_{d}\to\pi^+\pi^-$ 
decays. The left plot shows distributions for the lower
(dashed line) and higher (solid line) momentum pion in this decay. 
The right plot presents the later distribution in integral form,
which gives loss of efficiency as a function of the high momentum cut-off of
the particle ID device. \label{pid_fig_1}}
\end{figure}

In the design of any particle identification system, the dominant
consideration is the momentum range over which efficient separation of the
 charged hadron species $\pi$, $K$, and $p$ must be provided.
The upper end of the momentum
requirement is given by two-body decays such as 
$B\rightarrow \pi^{+}\pi^{-}$ which produce the highest momentum particles 
in $b$ decays.
Fig.~\ref{pid_fig_1} shows the momentum distribution of
pions from the decay $B_{d}^{o}\rightarrow \pi^{+}\pi^{-}$
 for the case where the two particles are within the spectrometer's
acceptance.

\begin{figure}[htb]
\vspace{-2mm}
\centerline{ \epsfxsize=2.5in \epsffile{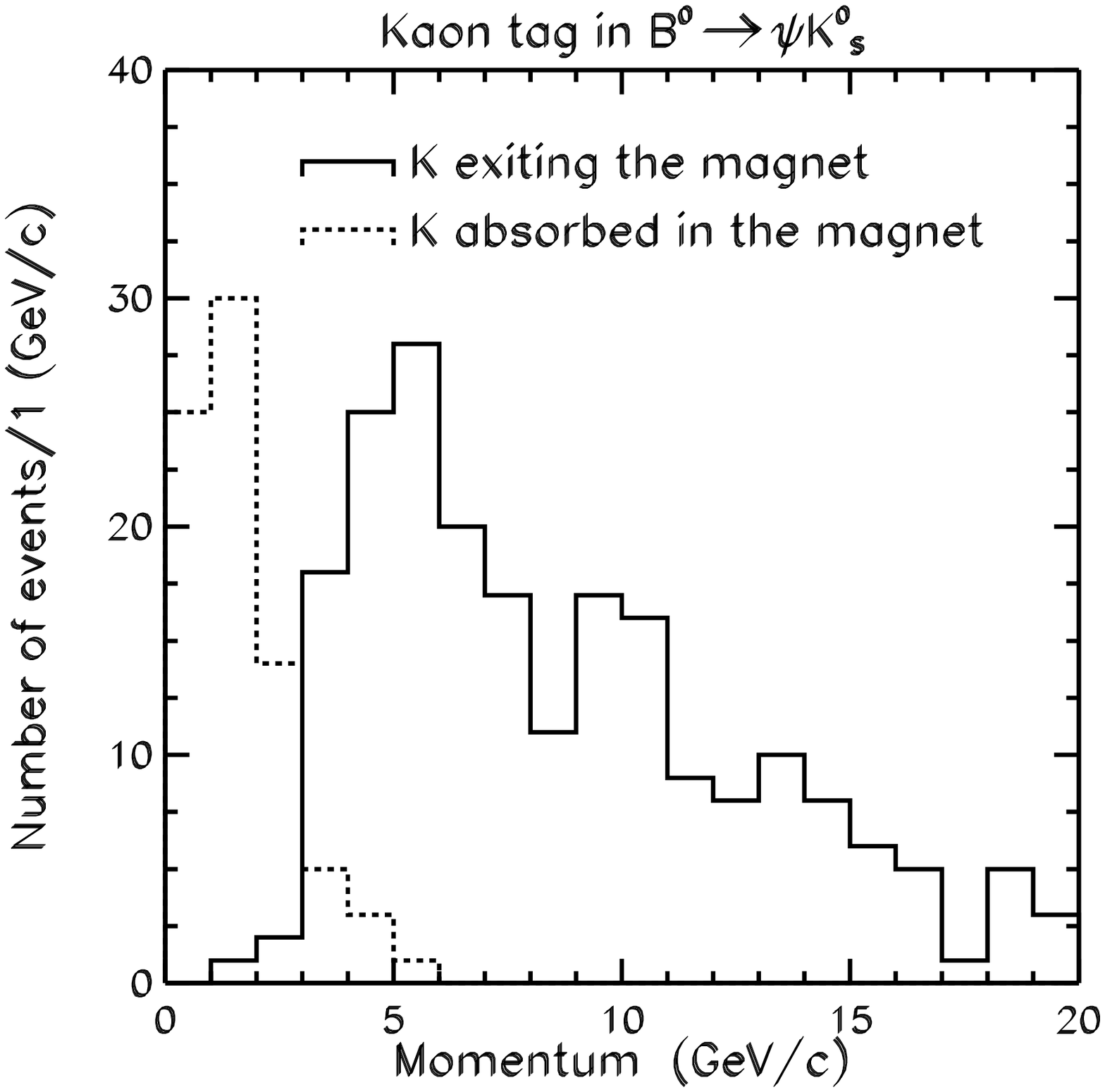}
             \epsfxsize=2.5in \epsffile{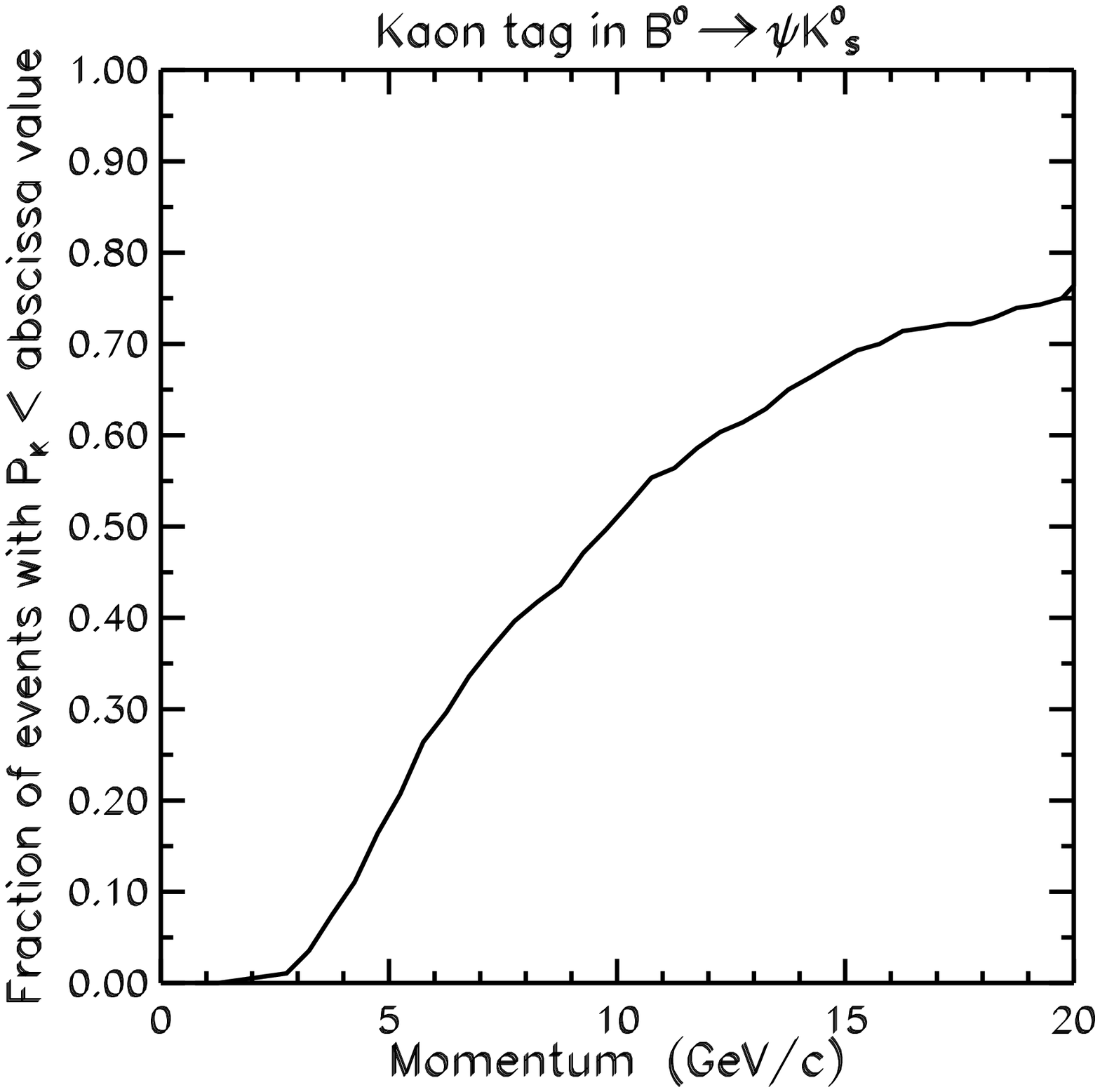}} 
\caption{The momentum distribution of `tagging' kaons for the case
where the signal particles ($\psi K^0_S$) are within the
geometric acceptance of the spectrometer.
The left plot shows distributions for kaons absorbed in 
(dashed line) and exiting from (solid line) the magnet.
The right plot presents the later distribution in integral form,
which gives loss of efficiency as a function of the low momentum cut-off of
the particle ID device. \label{pid_fig_2}}
\end{figure}

The low momentum requirement is defined by having high efficiency for `tagging'
kaons from generic B decays. Since these kaons come mainly from daughter
$D$ mesons in multibody final state $B$ decays, they typically have much lower
momentum than the particles in two-body decays.  Fig.~\ref{pid_fig_2} shows the
momentum distribution of  `tagging' kaons for the case where the signal
particles are within the geometric acceptance of the spectrometer. About 1/5 of
the tagging kaons never exit the end of the spectrometer dipole.  Almost all of
them are below 3 GeV, and most kaons exiting the dipole have momenta above 3
GeV. Based on these considerations, the momentum range requirement for the
particle identification  system is defined as separating kaons from pions
between 3 and 70 GeV/c.

\begin{figure}[htb]
\centerline{ \epsfxsize=5.0in \epsffile{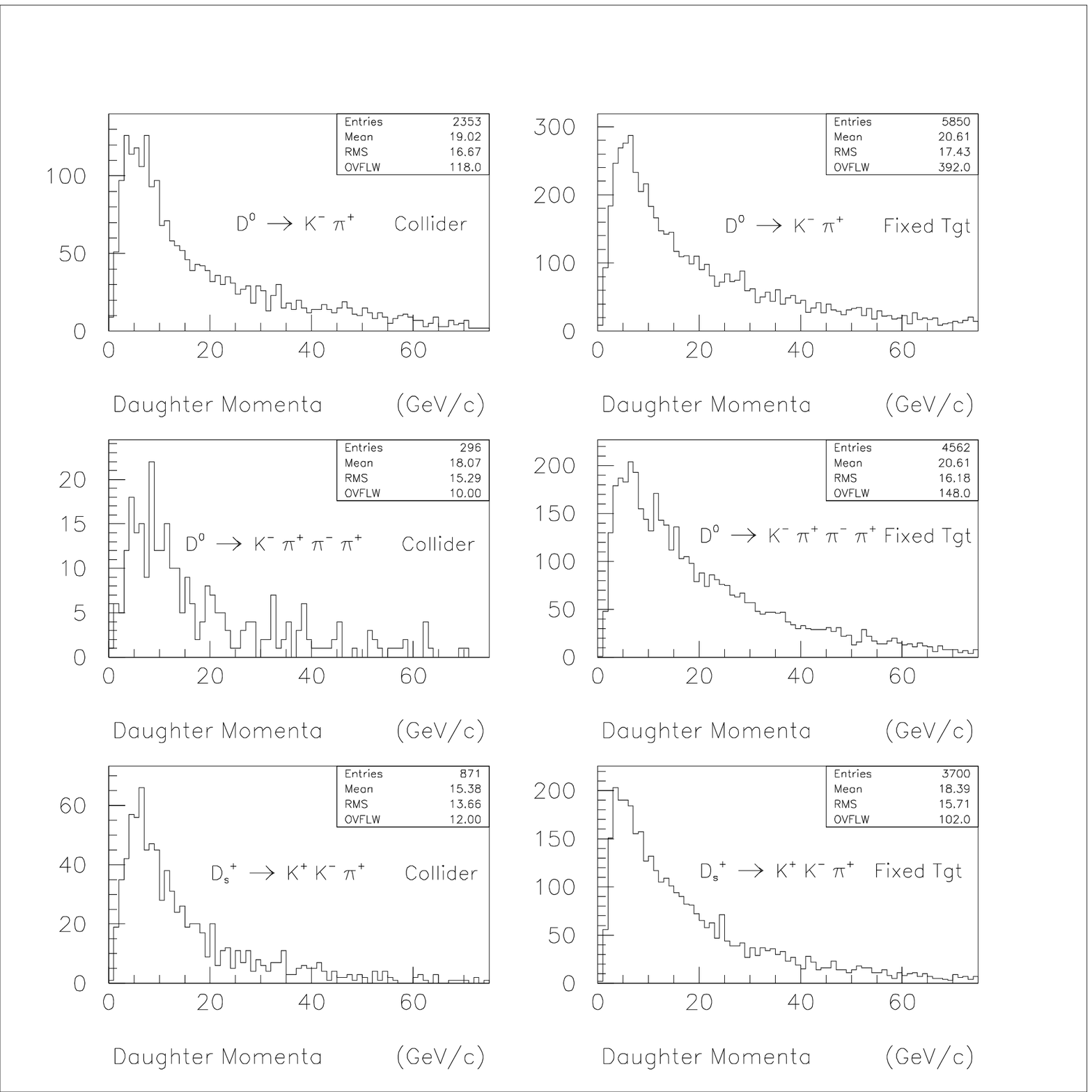}}
\caption{
The momentum spectra of kaons from accepted
$D^{o}\rightarrow K^{-}\pi^{+}$, $D^{o}\rightarrow K^{-}\pi^{+}\pi^{-}\pi^{+}$,
and $D_{s}^{+}\rightarrow K^{+}K^{-}\pi^{+}$ in both collider and fixed
target modes.
\label{pid_fig_3}}
\end{figure}

Kaons and pions from directly produced charm
decays have momenta which are not very different from the kaons from
$B$-decays. 
Fig.~\ref{pid_fig_3} shows the momentum spectra of kaons from accepted
$D^{o}\rightarrow K^{-}\pi^{+}$, $D^{o}\rightarrow K^{-}\pi^{+}\pi^{-}\pi^{+}$,
and $D_{s}^{+}\rightarrow K^{+}K^{-}\pi^{+}$ in both collider and fixed
target mode. The range set by the  $B$-physics requirements is a reasonable
 choice for charm physics.

Because of the large momentum range and  limited longitudinal space available
for  a particle  identification system in the C0 enclosure, the only sensible
of detector technology is a gaseous ring-imaging Cherenkov counter.
Fortunately, there are atmospheric pressure gas radiators which provide signal
separation between pions and kaons in this momentum region. 
Table~\ref{pid_tab_1} gives some parameters for two candidate radiator gases:
$C_{4}F_{10}$ and $C_{5}F_{12}$.  Note that below 9 GeV, these gases do not
provide $K/p$ separation since kaons are below threshold. $C_{5}F_{12}$ is used
in the barrel part of the DELPHI RICH and in SLD (mixed with 15\% N$_2$). It
needs to be operated at $40^o$C because of its high condensation point. The
$C_{4}F_{10}$ gas can be used at room temperature. It is used in the DELPHI
endcap RICH and was adopted for the HERA-B and LHC-B  RICH detectors.

\begin{table}
\begin{center}
\begin{narrowtabular}{1.5cm}{ccc} \hline 
Parameter       &  $C_{4}F_{10}$  &  $C_{5}F_{12}$   \\ \hline
(n-1) x $10^{6}$   &     1510        &      1750        \\
gamma-threshold &     18.2        &      16.9        \\
$\Theta_c$($\beta$=1) &     54.9 mr     &      59.1 mr     \\
$\pi$ threshold   &      2.5 GeV/c  &       2.4 GeV/c  \\
$K$ threshold     &      9.0 GeV/c  &       8.4 GeV/c  \\
$p$ threshold   &     17.1 GeV/c  &      15.9 GeV/c  \\ \hline 
\end{narrowtabular}
\caption{Parameters of the two candidate Cherenkov gas
         radiators. \label{pid_tab_1}}
\end{center}
\end{table}

The visible light photons can be detected either with a new
device called Hybrid Photo-diodes, HPD, or 
multi-anode PMTs such as the 16 channel tubes
from Hamamatsu used in the HERA-B detector \cite{Desalvo}.

In order to increase positive identification of low momentum particles,
one interesting possibility is to insert a thin ($\sim5$ cm) piece of aerogel
at the entrance to the gas RICH as proposed by LHC-B \cite{Forty}.
For example,
aerogel with refractive index of $n=1.03$ would lower the
$\pi$, $K$, $p$ momentum thresholds to
$0.6$, $2.0$, $3.8$ GeV/c respectively. 
Shorter wavelength Cherenkov photons undergo Raleigh scattering inside
the aerogel itself. They are absorbed in the radiator or exit at random
angles. A thin mylar or glass window between the aerogel and the gas radiator
would pass photons only in the visible range, eliminating the scattered
component. The same photodetection system could then detect Cherenkov rings
produced in both the gaseous and the aerogel radiators.
The radius of rings produced in the aerogel would be about 4.5 times 
larger than 
those produced in $C_{4}F_{10}$. 
The aerogel radiator would provide positive $\pi/K$ separation
up to 10-20 GeV/c. It would also 
close the lower momentum gap in $K/p$ separation.
Since the low momentum coverage would be provided by aerogel,
one could think about boosting the high momentum reach of the
gas radiator by switching to a lighter gas such as
$C_{2}F_{6}$ ($n=1.0008$) or $C\,F_{4}$ ($n=1.0005$).
This would also loosen the requirements for Cherenkov angle resolution
needed to reach a good $K/\pi$ separation at a momentum of 70 GeV/c.
Detailed simulations are needed to evaluate trade-offs due to more complicated
pattern recognition.

The alternative options to be considered for improving particle
identification at lower momenta include a time-of-flight system or a DIRC.

\subsubsection{Electron, photon and muon detection} BTeV is considering the use
of a cryrogenic liquid based EM calorimeter. One solution uses a detector
filled with liquid Krypton, similar to the one being used in NA48
\cite{krypton}. This would
provide the best possible position and energy resolution. It would in all
probability only be used if there were processes where photons or $\pi^o$'s
could be usefully reconstructed. A cheaper alternative, which is probably
adequate for electron identification is a lead-liquid Argon detector. The
liquid based detectors are naturally radiation hard.

\subsubsection{The detached vertex trigger}
BTeV places the vertex detector in the magnetic field to allow the elimination
of low momentum tracks in the trigger. These tracks have large multiple
scattering and can appear to come from detached verticies, thus fooling the
trigger.  

The goal of the trigger algorithm is to reconstruct  tracks and find vertices 
in every
interaction up to an interaction rate of order 10 MHz (luminosity of
$10^{32}$\,cm$^{-2}$\,s$^{-1}$ at $\sqrt{s} = 2$\,TeV). This entails an
enormous data rate coming from the detector ($\sim100\,$GB/s), thus a
careful organization of the data-flow is crucial.  
This  trigger must be capable of
reducing the event rate by a factor between a hundred and a thousand in the
first level.

The key ingredients for such a trigger are
a vertex detector with excellent spatial resolution, fast readout, and 
       low occupancy;
a heavily pipelined and parallel processing architecture well suited 
        to  tracking and vertex  finding; 
inexpensive processing nodes, optimized for specific 
      tasks within the overall architecture;  
 sufficient memory to buffer the event data while the calculations 
      are carried out; and 
 a switching and control network to orchestrate the data movement
      through the system.

The detailed reference trigger scheme proposed for BTeV was developed by the 
Univ. of Pennsylvania Group \cite{selove}. 
While it may undergo some revisions, it has been an excellent starting
point for studying this kind of trigger and the initial results for this
particular algorithm are very encouraging.

\begin{figure}
\vspace{-2.5cm}
\centerline{\epsfxsize=8.0in  \epsffile{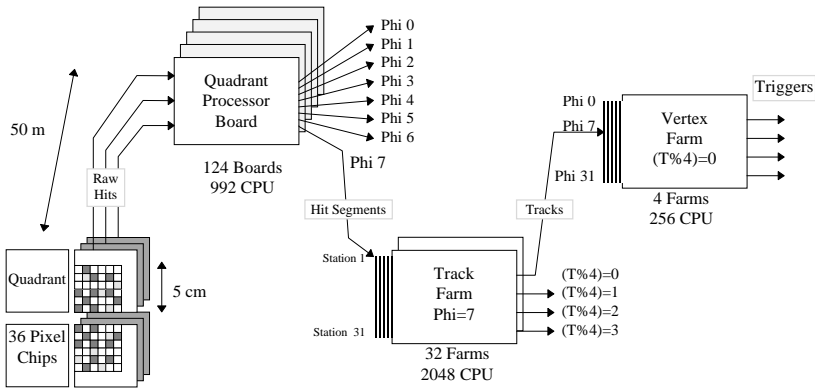}}
\vspace{-5cm}
\caption{Schematic of Proposed  Level I vertex trigger
         \label{trig_fig_1} }
\end{figure}

The proposed algorithm has four steps (see Fig.~\ref{trig_fig_1}): 

In the first step, hits from each pixel plane are assigned to detector
subunits.  It is desirable to subdivide 
the area of the detectors 
in a way which minimizes the number of tracks crossing from one 
subunit into the next and assures uniform and low occupancy ($<1$ hit/event) 
in all subunits. Therefore each plane is divided into 32 azimuthal 
sectors (``$\phi$-slices").
Simulation shows that at 2 TeV, on average each $\phi$-slice contains 0.2 hits
per inelastic interaction, approximately independent of the plane's $z$
coordinate.

In the second step, ``station hits" are formed in each triplet of 
closely-spaced planes using hits from 
each $\phi$-slice. Since the pixels under
consideration have a rectangular shape (e.g.\ 30\,$\mu$m by 300\,$\mu$m), the 
two outer planes in each station are 
oriented for best resolution in the bend 
view, and the inner plane for the non-bend view. There is a ``hit
processor" 
for each $\phi$-slice of each station.
The hit processor finds triplets of overlapping pixels, and sends to the
next stage a single ``minivector"
consisting of $x$, $y$, $\frac{dx}{dz}$ and $\frac{dy}{dz}$. 
Given the good position
resolution of each measurement, a detector station determines a
space point accurate to $5-9\,\mu$m in $x$ and $y$, and $y$
and $x$ slopes to $\frac{dy}{dz}\approx
1\,$mr (bend-view) and $\frac{dx}{dz}\approx 10\,$mr (non-bend).
The  use of these minivectors in the track-finding stage substantially
reduces combinatorics.

In the third step, minivectors in each $\phi$-slice from the full set of
stations are passed via a sorting
switch to a farm of ``track processors." The sign of the bend-view slope is used
to distinguish forward-going and backward-going minivectors and send them to
separate farms.
To handle the few percent of tracks crossing segment
boundaries, hits near a boundary between two $\phi$-slices can be 
sent to the processors for both slices. 
Each track processor links minivectors into tracks, proceeding along $z$
from station to 
station. For each pair of minivectors in adjacent stations, 
the processor averages the $y$-slopes in the two stations and 
      uses this average slope
      (which represents the slope of the chord of the magnetically
      deflected trajectory) to project from the first station into the
      second. It then checks whether the $y$-value of the
      minivector in the downstream station agrees with the projection
within an acceptance window. 
If three or more hits satisfy this requirement, a fast fitting algorithm
      finds the momentum of the track
      candidate.

In the fourth step, the tracks are passed to a farm of ``vertex 
processors" and used to form vertices.
The vertex with $\sum{ p_z}$ closest to zero
is designated as the primary vertex. Tracks which have
a large impact parameter with respect to this vertex 
are taken as evidence for heavy quark production in
an event. 
To reduce the effect of multiple scattering on vertex resolution,
tracks below an
adjustable bend-view transverse-momentum ($p_y$) threshold are excluded from the
vertex finding.

\subsubsection{Beyond the Level I Trigger}

\begin{figure}[t]
\centerline{ \hbox{
\epsfxsize=4.0in \epsffile{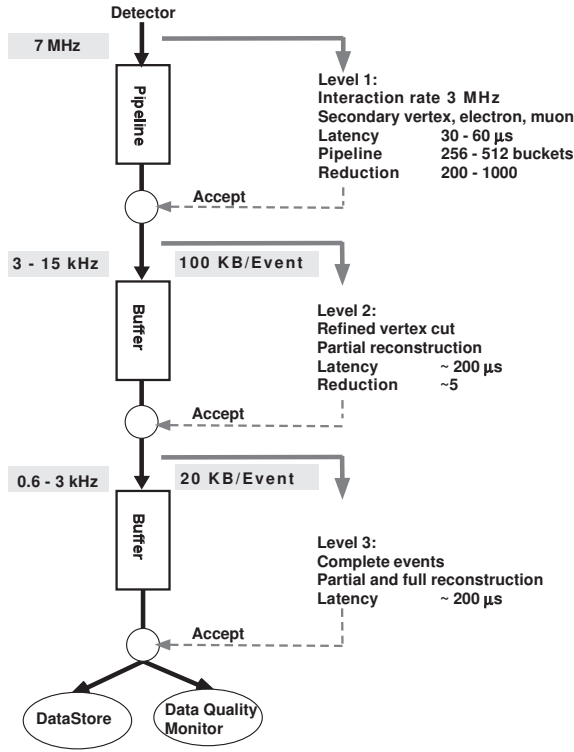}
}}
\caption{ The BTeV data flow diagram. The numbers are for the nominal luminosity
of $\rm 5 \times 10^{31} cm^{-2}s^{-1}$.
         \label{trig_fig_2} }
\end{figure}

Modern experiments in High Energy physics implement hierarchical trigger systems
and BTeV is no exception.   Many details of the full data acquisition system have
to be worked out as the design of the detector components and trigger algorithms
become more mature.   At the design
luminosity of $\rm 5 \times 10^{31} cm^{-2}s^{-1}$ and with a total  inelastic
cross section of $\sigma_{inel} \approx 60$ mb the interaction rate at BTeV
will be around 3 MHz. Fig.~\ref{trig_fig_2} gives the outline of the expected
data flow at the nominal design luminosity. 

The Level I trigger  must be capable of reducing the trigger rate  by a
factor  between a hundred and a thousand. 
This seems to be achievable even if the only test made is for a detached vertex
in vacuum.  The event rate is then reduced  to  $\approx$100~kHz in the worst case
during the highest luminosity running.  It is still possible to move all this
data, amounting to a maximum rate of 10~GB/s, to the Level II trigger.  

Events that pass the first trigger level are forwarded to compute nodes to be
further analyzed.  First, detector component information is used to  partially
reconstruct the event, e.g. finding tracks in the vertex detector. Algorithms with
refined secondary vertex requirements or invariant mass cuts will be implemented at
this level.  Events remaining after this step are then fully assembled in a
switch-based eventbuilder and passed to a processor farm.  Alternatively, an
algorithm such as associating a mass with each detached vertex can be computed at
Level I which would eliminate the need for an intermediate trigger step before the
eventbuilder.  Since most of the detached-vertex events will be $K_{s}$, a
requirement that the mass at the vertex be above some threshold  (say
1~GeV/c$^{2}$) could reduce the rate to less than 10~kHz with almost no bias
against charm and beauty events of interest.  Even so, it will probably be
impractical to  write several kHz of events to archival storage.

The final trigger level will combine information from different
sub-detectors to further suppress background events.  
Particle identification will be available at this stage and could
be used to obtain a very clean direct charm signal for specific final states.

Because the event rate surviving  this last level may still be close to a kHz,
the software will  probably have to  reduce the amount of data per event to
archival  storage by writing out an event summary which eliminates much of the
raw  data.  The event summary would be around 20~KB so that the output rate 
could still be as high as 20~MB/s. This results in a dataset size of 200~TB/yr,
comparable to what is expected from CDF or D0 in Run II.  However, initial
phases of the experiment will run at  much lower rates and  will be comparable
in dataset size to a current fixed target experiment.

\subsubsection{Trigger simulations}

Track hits are generated with multiple scattering and energy loss properly
taken into account. The beam width is taken as $50~\mu m$ r.m.s., in both coordinates
travsverse to the beam and the interaction region length is taken as 30 cm
r.m.s.
 The reconstruction program does the full pattern recognition
starting with hits on the pixel planes. Hits are formed into tracks, using the
$\phi$ slices as discussed above, and the track momenta are calculated. The
algorithm allows rejection of all tracks whose momentum in the bend plane,
$p_y$, is less than a fixed cut. All the results shown here are for $p_y > 0.5$
GeV/c, though this cut has not been optimized.

This algorithm is preliminary and can be imporved. The current
simulation   does not take into account Moli\`{e}re multiple-scattering tails,
the effect of pair-conversions or hadronic interactions in the silicon,
detector inefficiencies, or noise in the detector.  The trigger performance has
only been studied when there is one interaction per crossing. There are several
ideas on how to handle the situation with more than one interaction.

\begin{figure}
\epsfig{figure=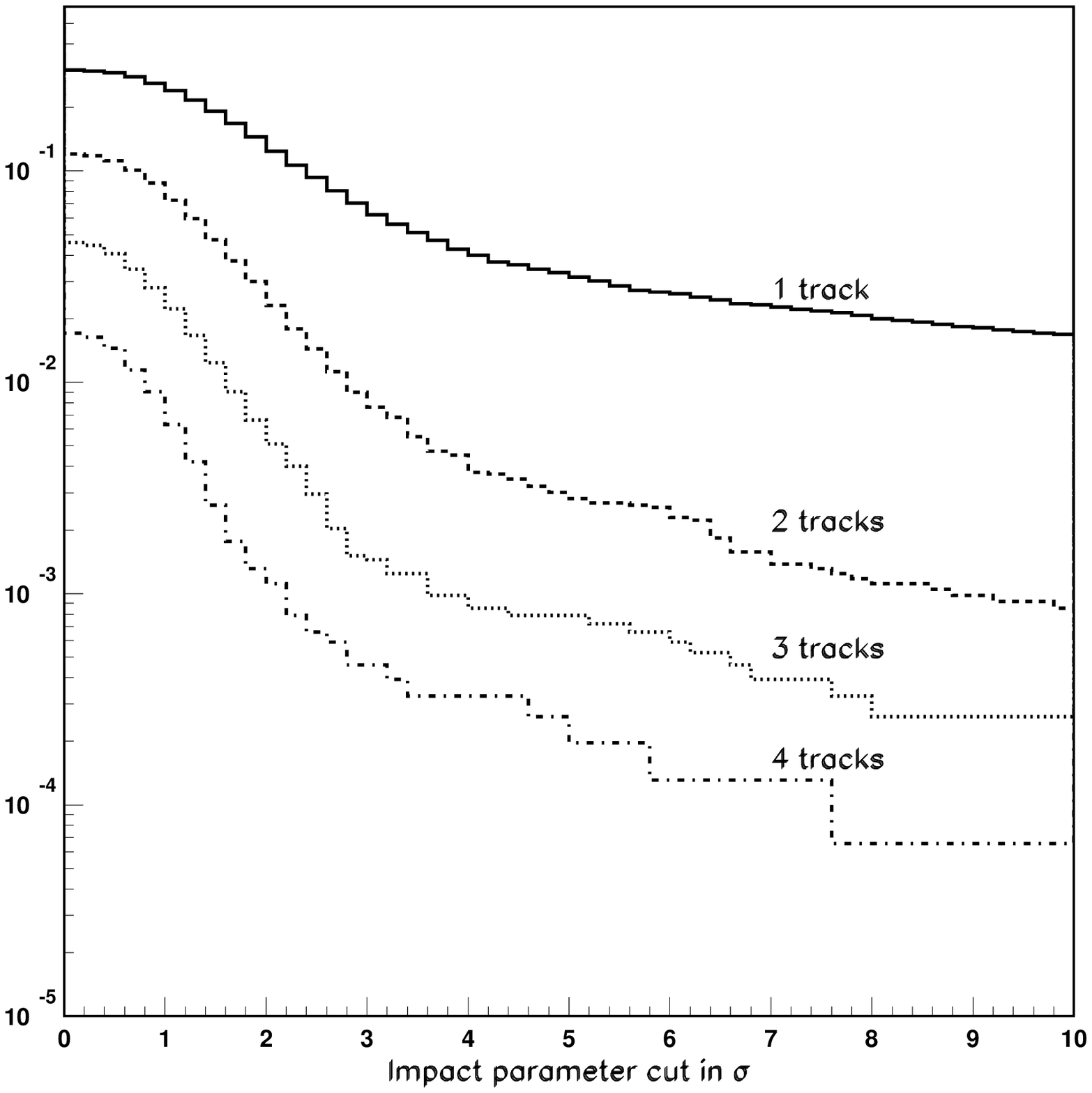,width=2.5in}
\epsfig{figure=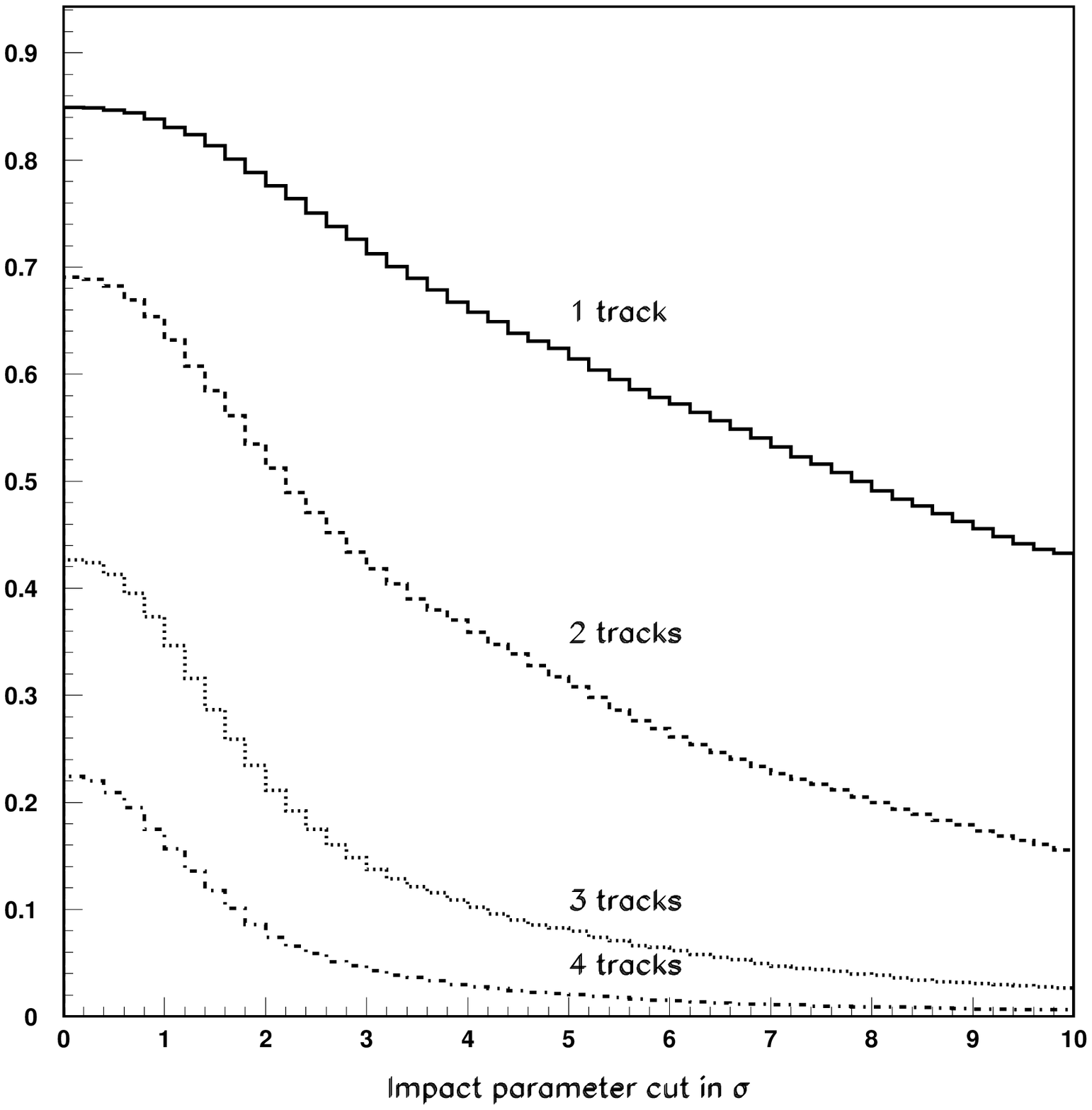,width=2.5in}
\caption{\label{uds_trig_eff}Trigger efficiency of light quark background
(left) and for $B^o\to\pi^+\pi^-$ events (right), both shown on the ordinate.
The efficiency for the $\pi^+\pi^-$ plot is for events with tracks both
required to be in the solid angle of the detector.
The abscissa gives the value of the impact parameter in terms of number of standard deviations ($\sigma$) 
of the track from the primary vertex. The curves show the effect of requiring
 different numbers of tracks. All tracks are required to have at least
 0.5 GeV/c momentum in the bend plane.}
\end{figure}

The trigger algorithm first finds a primary interaction vertex. The distance a
track misses the primary is then determined in units of $n\sigma$, where
$\sigma$ refers to the r.m.s. error for a particular track distance to the
primary. Then the number of tracks which miss the primary by $n\sigma$ is
determined. Fig~\ref{uds_trig_eff} shows the efficiency for triggering light
quark ($uds$) background events as a function of both the number of tracks
required and the specfic $n\sigma$ cut. The efficiency to trigger on a $B^o\to
\pi^+\pi^-$ is also shown for events with both tracks in the detector.

When the experiment runs one value of $n\sigma$ and one value of the number of
tracks needs to be chosen. Values of 3$\sigma$ and 2 tracks provides an
efficiency, in the spectrometer, of 43\% for this process with a rejection of
125:1 against light quark background. The final cut values must be chosen with
reference to many processes, however.

\begin{figure}
\begin{center}
\epsfig{figure=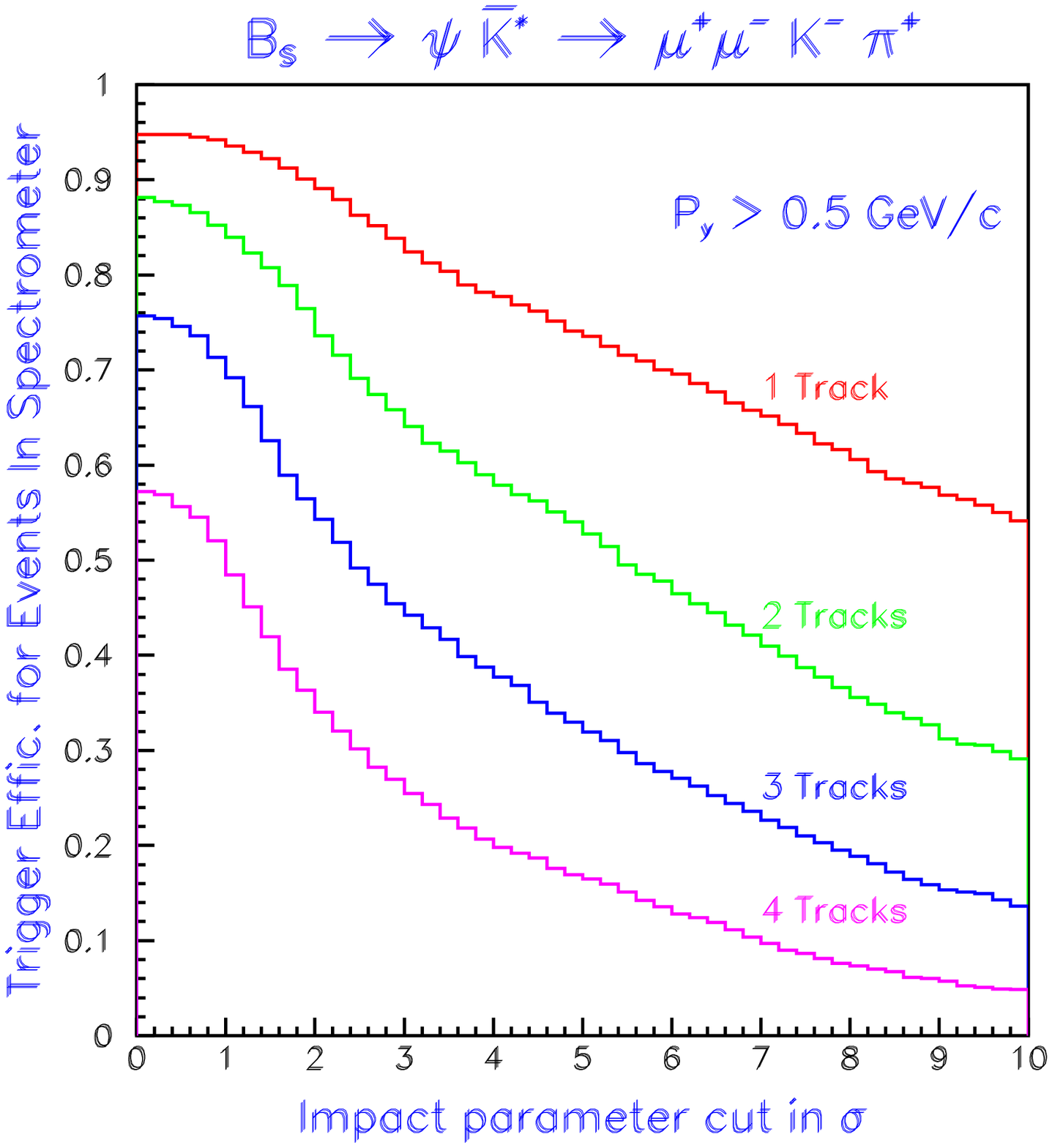,width=2.6in}
\epsfig{figure=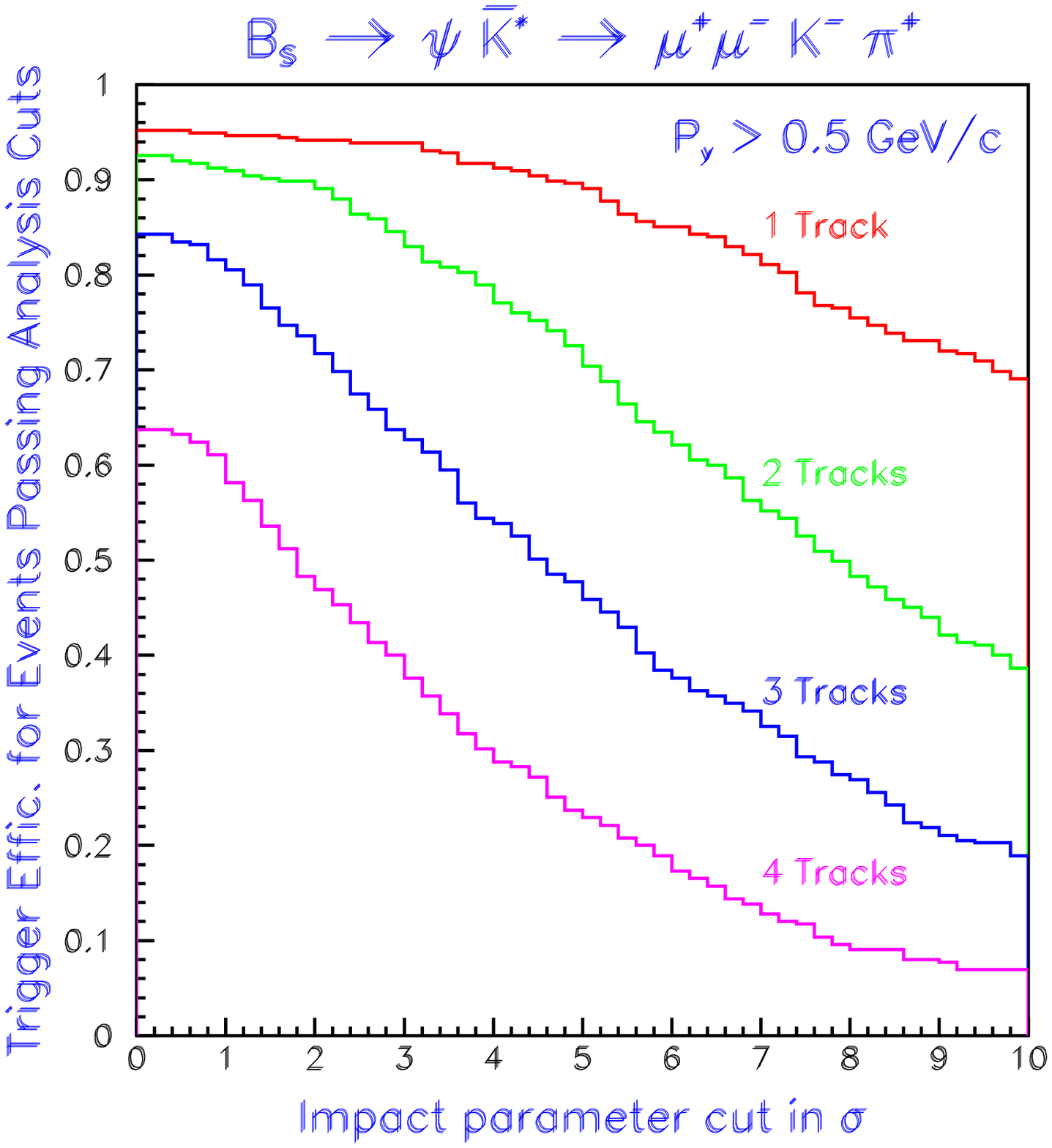,width=2.6in}
\end{center}
\caption{\label{psiks_trig}Trigger efficiency of $B_s\to\psi
K^{*o}$, $\psi\to\mu^+\mu^-$, $K^{*o}\to K^-\pi^+$ for tracks in the geometric
acceptance(left) and after all analysis cuts (right).
The abscissa gives the value of the impact parameter in terms of number of standard deviations ($\sigma$) 
of the track from the primary vertex. The curves show the effect of requiring
 different numbers of tracks. All tracks are required to have at least
 0.5 GeV/c momentum in the bend plane.}
\end{figure}

The trigger efficiency on states of interest is correlated with the analysis
criteria used to reject background. These criteria generally are focussed upon
insuring that that the $B$ decay track candidates come from a detached vertex,
that the momentum vector from the $B$ would point back to the primary
interaction vertex, and that there aren't any other tracks which are consistent
with $B$ vertex. When these criteria are applied first and the trigger
efficiency estimated after, the trigger efficiency appears to be larger.
In Fig.~\ref{psiks_trig} I show the efficiency to trigger on $B_s\to\psi
K^{*o}$, $\psi\to\mu^+\mu^-$, $K^{*o}\to K^-\pi^+$ using the tracking trigger
only for events with the four tracks in the geometric acceptance, and the
efficiency evaluated after all the analysis cuts have been applied.
Here the trigger efficiencies for $3\sigma$ and 2 tracks are 67\% for events
with all 4 tracks in the geometrical acceptance and 84\% on events after all
the analysis cuts have been applied.

\subsection{The LHC-B design}

\begin{figure}[htb]
\centerline{\epsfig{figure=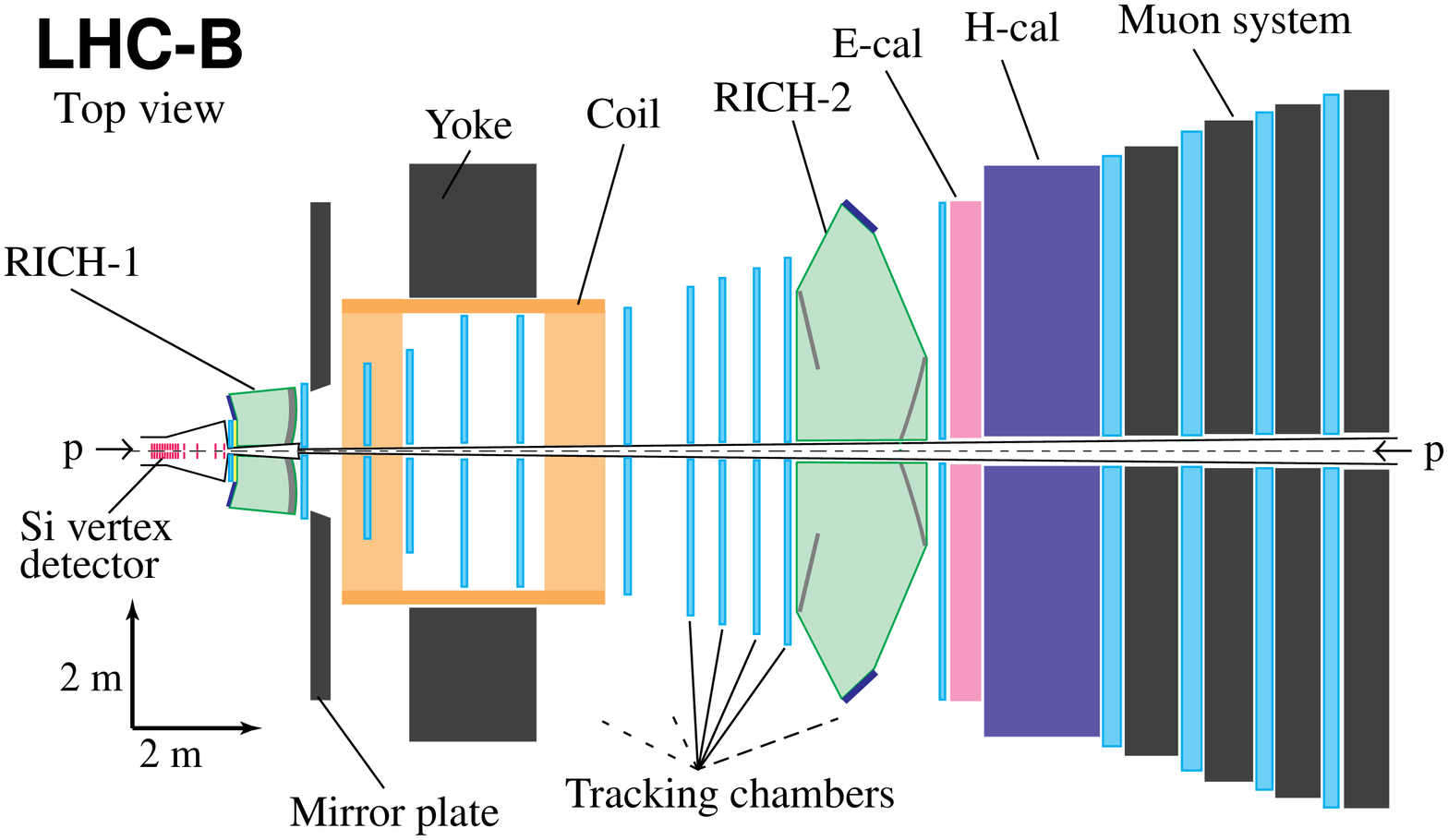,width=5.4in}}
\caption{\label{LHCB_det} Overview of the LHC-B detector.}
\end{figure}
I will not give as an extensive discussion of the LHC-B design as I have done
for BTeV, as the fundamental design concepts are quite similar. The LHC-B
detector schematic is shown in Fig.~\ref{LHCB_det}.

There are several significant differences between LHC-B and BTeV. LHC-B is a
single arm device. It also includes a hadron calorimeter, which BTeV lacks.
The hadron calorimeter is used in the trigger for rejecting events with multiple primary
interactions. The interaction region length is much smaller at LHC than at the
Tevatron, so the vertex detector can be smaller. However, the LHC-B vertex
detector is outside the magnetic field. Furthermore, since there is a RICH
detector between the vertex detector and the magnetic, most of the momentum
information comes from the tracking chambers alone. The vertex detector current
is designed as silcon strips. There is a second RICH detector to indentify the
higher momentum tracks. 

\subsubsection{The vertex detector} 
\begin{figure}[htb]
\centerline{\epsfig{figure=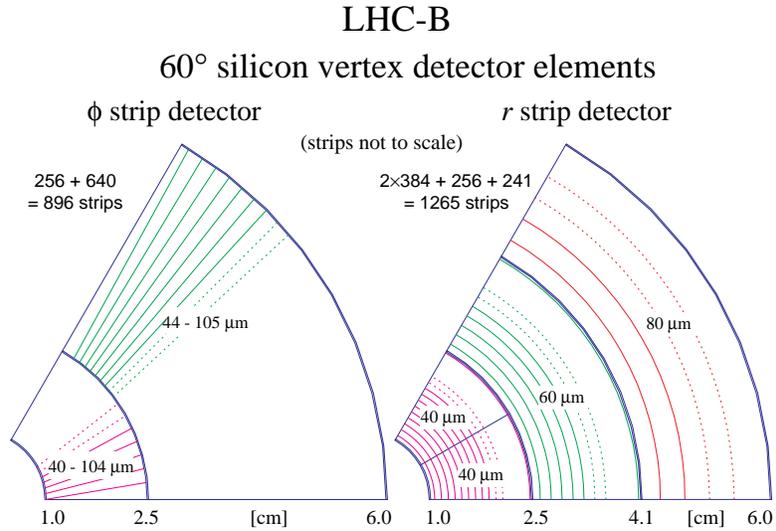,width=4in}}
\caption{\label{LHCB_vd} Overview of the LHC-B vertex detector.}
\end{figure}

\begin{figure}[htb]
\centerline{\epsfig{figure=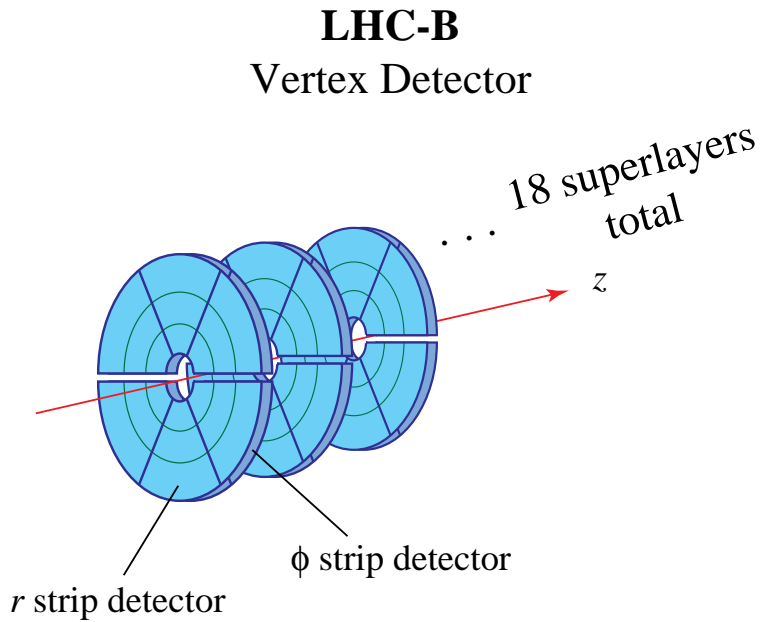,width=4in}}
\caption{\label{LHCB_silicon} Details of the LHC-B silicon strip vertex detector.}
\end{figure}

The LHC-B vertex detector contains 18
superlayers containing one $r$ strip detector and one $\phi$ strip detector.
The arangement of superlayers is shown in Fig.~\ref{LHCB_vd}. The strip
arrangements on the individual detectors is shown in Fig.~\ref{LHCB_silicon}.
The segmentation is increased closer to the beam line because the occupancies
are higher there. 


\subsubsection{The RICH detectors}

\begin{figure}
\centerline{\epsfig{figure=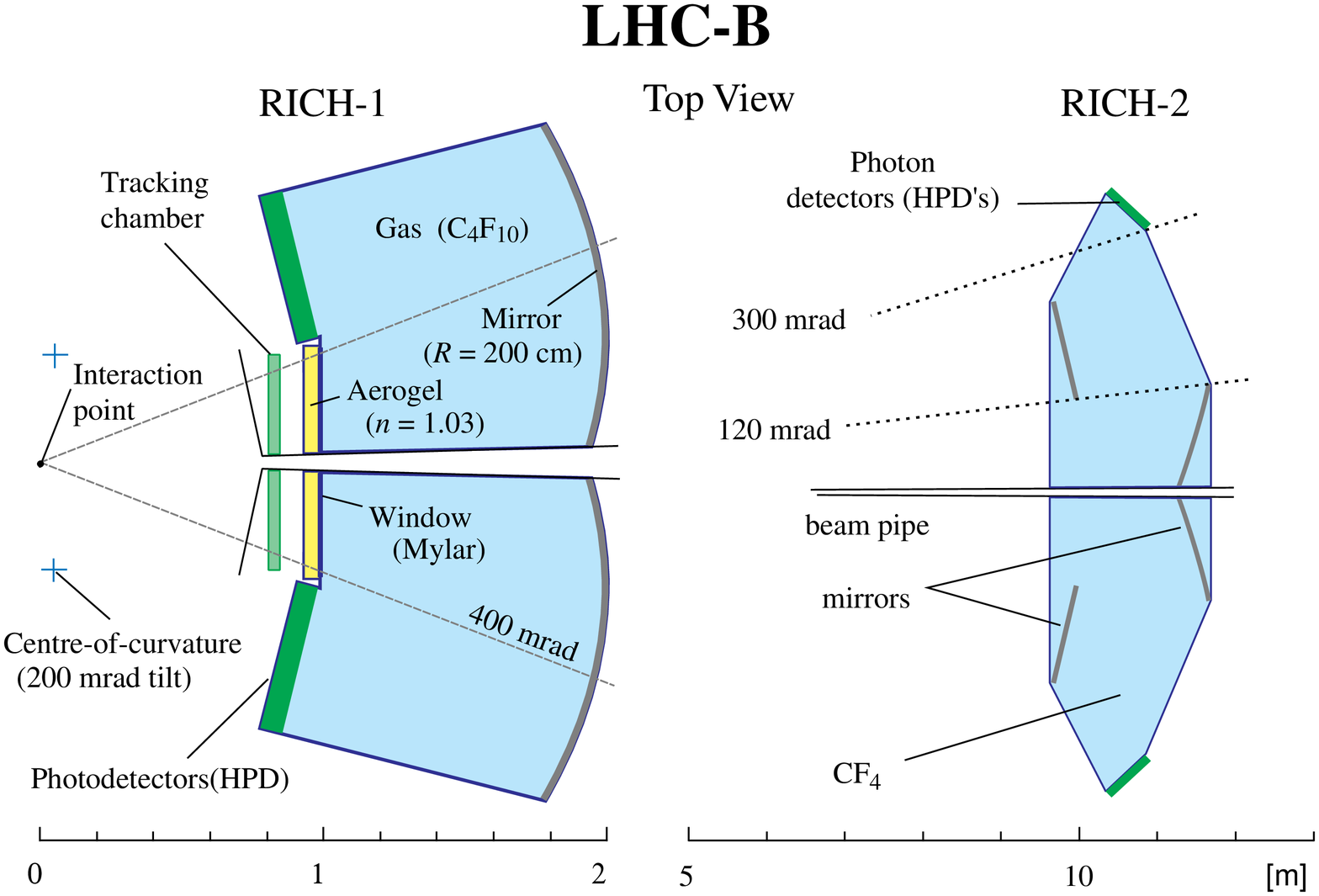,width=4in}}
\caption{\label{LHCB_rich} Schematic of the LHC-B RICH detectors. The scale on
the bottom gives the location with respect to the center of the interaction
region.}
\end{figure}

LHC-B has two RICH detectors to cover the large momentum range. They are shown
in Fig.~\ref{LHCB_rich}. The both use HPD's as photon detectors.
The first one,
located before the magnet and after the silicon detector has large acceptance
for charged hadrons. It uses a C$_4$F$_{10}$ gas radiator and possibly an
aerogel radiator. The light is mirrored to photon detectors out of the
acceptance. The second RICH uses a lower index gas, CF$_4$ to identify higher
momentum hadrons, more concentrated at smaller angles with respect to the beam.
There is a double mirror system to keep the photon detectors out of the 
overall acceptance.  

\subsubsection{Calorimeters and Muon detector}

The electromagnetic calorimeter has not been completely specified. One
possibility is to have an inner part made of radiation hard PbWO$_4$ crystals,
with the outer part be a a lead-scintillator ``Shaslik" system. A ``preshower"
detector is also envisaged whos function is to reject photons in the electron
trigger.

The hadron calorimeter also could have an inner and outer dicotomy. The inner
part which needs high radiation resistance may possibly be constructed with
tungsten plates and quartz as the active material generating Cherenkov light.
The outer part could be alternating plates of steel and plastic scintillator.

The Muon detector consists of three 70 cm thick iron slabs with tracking
chambers placed downstream of the hadron calorimeter.

\section{Simulations}

\subsection{Introduction}
I will give the results of several simulations done for BTeV. There are LHC-B
simulations which can be found in their Letter of Intent \cite{LHCBintent}.
 
 BTeV has developed several fast simulation packages to verify the basic
concepts and aid in the final design. Collisions are generated using either
Pythia 5.7 or Jetset 7.4. Beauty and Charm decays are modeled using the CLEO
decay generator QQ. Charged tracks are generated and traced through different
material volumes including detector resolution, multiple scattering and
efficiency. This allows measurement of  acceptances and resolutions in a fast
reliable manner. Pattern recognition has not been implemented. The key program 
is called MCFast \cite{MCFast}.

\subsection{Flavor tagging}
\subsubsection{Introduction}
In order to measure CP violation or mixing in a neutral $B$ decay, it is
essential that it be know if the intial particle was a $B$ or $\overline{B}$.
Since hadrons containing $b$ quarks are produced in pairs in $e^+e^-$ and
$p\overline{p}$ collisions, and thought to be in $pp$ interactions, one way to
accomplish this ``tagging" is to measure the $b$ content (flavor) of the other
produced $B$. It is easiest to view the problem in the $p\overline{p}$
situation. Here if a $\overline{B}^o$ is produced the acompanying particle is
either a $B^+$, $\Lambda_b$ or other baryonic state, $B^o$, or $B_s$.  Since
the $B^+$ doesn't mix, its flavor at production is same as when it decays. The
neutrals $B$'s, however, do mix and this dilutes the purity of the information.
In fact, about 20\% of the $B^o$'s turn up as $\overline{B}^o$ and  50\% of the
$B_s$'s turn up as $\overline{B}_s$. In $e^+e^-$ annihilations at the
$\Upsilon$(4S), only $B^o$ is produced in conjunction with $\overline{B}^o$,
and there are quantum correlation effects which need to be taken into
consideration to understand the dilution due to mixing \cite{BigiSanda}.

Another method of tagging is to use information from the fragmentation of the 
jet which produced the $B^o$. The clearest example of this occurs in charm. A
$D^{*+}$ is produced which decays into $\pi^+D^o$. The $\pi^+$ tags the flavor
of the $D^o$. A $\overline{D}^o$ would be tagged by a $\pi^-$. Unfortunately in
the $B$ case, the $B^*$ is not massive enough to decay into a charged pion and
decays only via a photon. There is however a $B^{**}$ state which can decay
into a charged pion and a $B^o$. It is also possible to try and find the charge
of pion closest in phase space to the $B^o$. These techniques are called ``same
side tagging."

Beside mixing effects, dilution will occur because of incorrect tagging
proceedures. The following definitions will be useful:
\begin{itemize}
\item $N\equiv$ number of reconstructed signal events we wish to tag.
\item $N_R \equiv$ number of right sign flavor tags.
\item $N_W \equiv$ number of wrong sign flavor tags.
\item $\epsilon = \left(N_R+N_W\right)/N$, this is the efficiency.
\item $D=(N_R-N_W)/(N_R+N_W),$ this is the dilution.
\end{itemize}

The quantity that enters directly in calculating the error on any particular
asymmetry measurement is $\epsilon D^2$.

\subsubsection{Kaon tags}
\begin{figure}[htb]
\vspace{-1.3cm}
\epsfig{figure=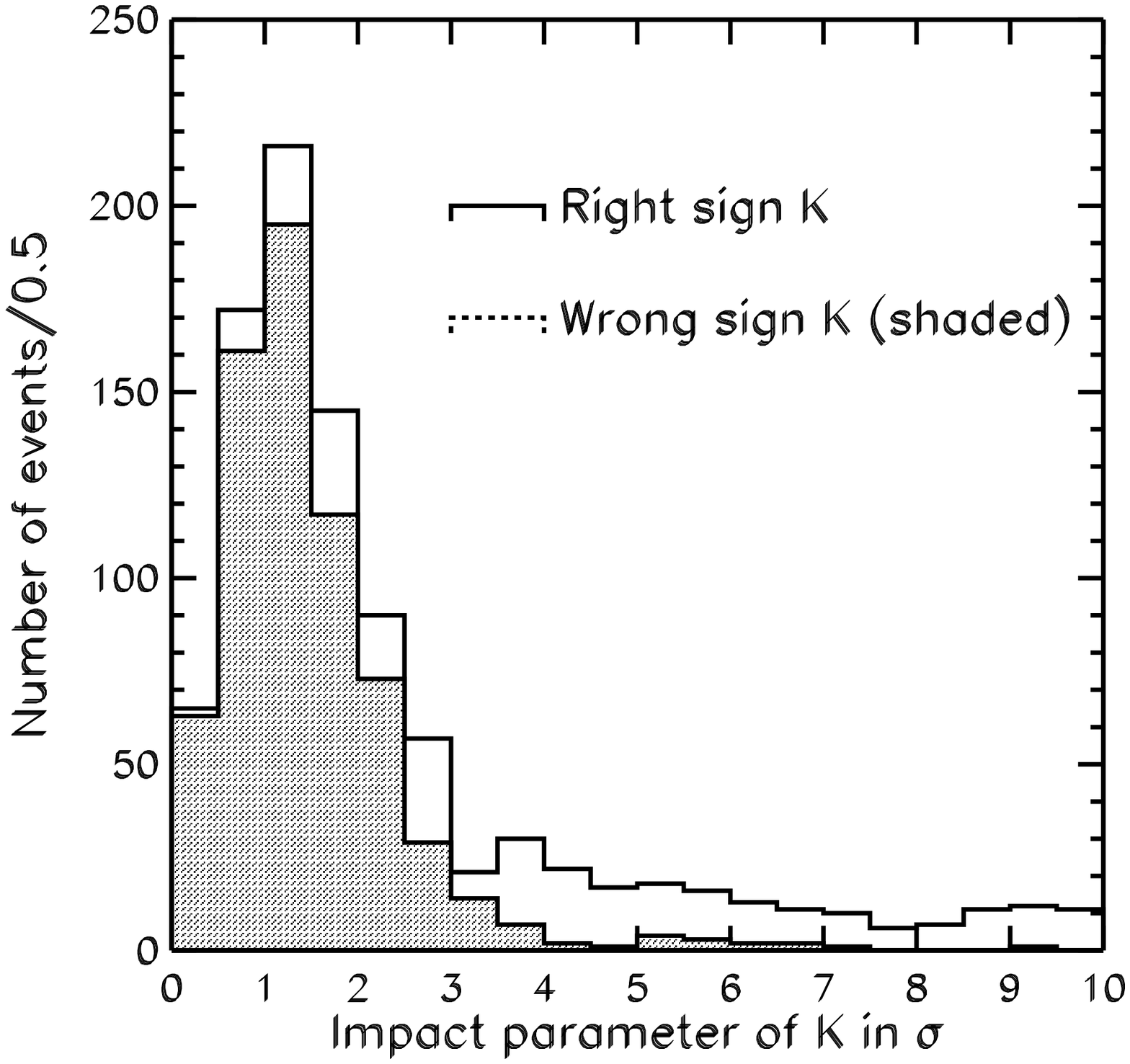,height=2.8in}\
\epsfig{figure=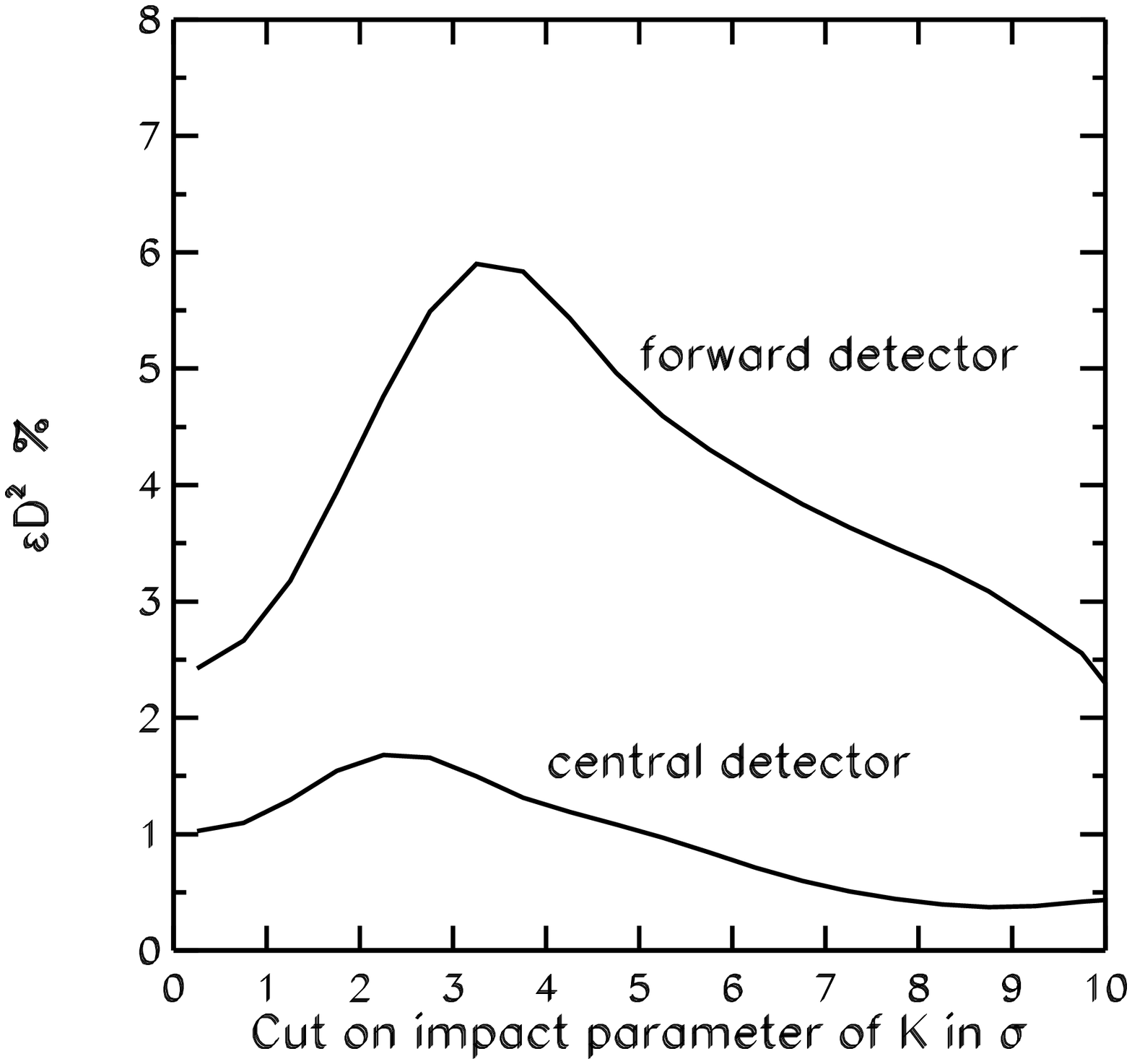,height=2.8in}
\caption{\label{ipk}(left) $L/\sigma$ distributions in BTeV for $K^{\pm}$ impact
parameters for right sign (unshaded) and wrong sign (shaded) tags. (right) Overall $\epsilon$D$^2$ values from kaon tagging for a forward detector
containing a flourine based RICH versus a central detector with 100 ns time of
flight resolution as a function of kaon impact parameter in units of
$L/\sigma$.
(Protons have been ignored in both cases.)}
\end{figure}

The decay chain $\overline{B}\to D\to K^-$ can be used to tag flavor.
BTeV has investigated the feasiblity of tagging kaons using a gas Ring
Imaging Cherenkov Counter (RICH) in a forward geometry and compared it with
what is possible in a central geometry using Time-of-Flight counters with
good, 100 ps, resolution. For the forward detector the momentum coverage
required is between 3 and 70 GeV/c.
The momentum range is much lower in the central detector
but does have a long tail out to about 5 GeV/c.  
Fig.~\ref{ipk} shows the number of identified kaons plotted versus their
impact parameter divided by the error in the impact parameter for both right
sign and wrong sign kaons. A right sign kaon is a kaon which properly tags the
flavor of the other $B$ at production. We expect some wrong sign kaons from
mixing and charm decays. Many others just come from the primary. A cut 
on the impact parameter standard deviation plot at $3.5\sigma$ gives an overall
$\epsilon D^2$ of 6\%. This may be an over-estimate because proton fakes
can be in the sample. Putting these in lowers $\epsilon D^2$ to 5.1\%.
These numbers are for a perfect RICH system. Putting in a fake rate of several
percent, however, does not significantly change this number.

The simulation of the central detector gives much poorer numbers. In
Fig.~\ref{ipk} (right) $\epsilon D^2$ for both the forward and central detectors are
shown as a function of the kaon impact parameter (protons have been ignored).
It is difficult to get $\epsilon D^2$ of more than 1.5\% in the central
detector. 

This analysis shows the importance of detecting protons. Both LHC-B
and BTeV are in the process of seeing if low momentum K/p separation can be
obtained using another radiator.

\subsubsection{Lepton tags}
The semileptonic decay process ${B}\to X \ell^+{\nu}$ provides, in principle,
a excellent way of tagging. The branching ratio to either muons or electrons is
large, $\sim$10\%. Unfortunately, the process $B\to\overline{D}X', 
\overline{D}\to Y\ell^-\bar{\nu}$ produces opposite sign leptons and is hard to
separate at hadron colliders. It may be possible to distinguish the 
charm decay vertex from the $b$ decay vertex in a forward detector, but this
has not yet been investigated. Not using this idea, $\epsilon D^2$ for muon is
1.5\% in the forward detector and 1.0\% in the central detector.
Electrons are similar, but somewhat worse giving $\epsilon D^2$ of 1\% in the
forward detector.

\subsubsection{Jet charge tags}
Jet charge is a weighted measure of the charge of the jet producing the tagging
$b$. This technique was invented at LEP. It has been successfully used by CDF
\cite{CDFjet}.
They define the jet charge as
\begin{equation}
Q_{jet} ={{\Sigma q_i\left|\overrightarrow{p_i}\cdot \overrightarrow{a}\right|}
\over{\Sigma\left|\overrightarrow{p_i}\cdot \overrightarrow{a}\right|}} ,
\end{equation}
where the sum is over all particles in the jet cone and $\overrightarrow{a}$ is
the jet axis.

\subsubsection{Summary of tagging efficiencies}

CDF has used their data to measure the tagging efficiencies they now achieve
and to project what the expect after detector improvements which are either in
progress or which they may propose. Currently CDF does not have any kaon
tagging. However, they believe a time-of-flight system would allow them to
achieve an $\epsilon D^2$ of 3\%. BTeV simulations indicate that only 1.5\% is
possible. Currently CDF measures a same side tagging (SST) efficiency of
$(1.5\pm 0.4)$\%. They extrapolate to 2\% because their new silicon vertex
detector will give a cleaner selection of fragmentation tracks. Using jet
charge they currently measure ($1.0\pm 0.3$)\% and extrapolate to 3\%, again
due the improved silicon detector.

The tagging efficiency projections are summarized for BTeV and a Central
Detector in Table~\ref{tab:tags}. BTeV expects, without proof, that their superior
pixel vertex detector will allow even better efficiencies than CDF, but for now
uses the CDF projected values.
For BTeV $\epsilon D^2$ is projected to be larger than 10\%. In subsequent
simulations 10\% will be used.

\begin{table}[t]
\caption{Projected tagging efficiencies ($\epsilon D^2$).\label{tab:tags}}
\vspace{0.4cm}
\begin{center}
\begin{tabular}{lcccccc}   \hline
 & kaon & muon & electron & SST & jet charge & sum\\\hline
BTeV &  5\% & 1.5\% & 1.0\% & $>$2\% & $>$ 3\% & $>$10\%\\
Central Detector & 0\% & 1.0\% & 0.7\% & 2\% & 3\% & 6\%
 \\ \hline
\end{tabular}
\end{center}
\end{table}

\subsection{Measurement of $x_s$}

BTeV has studied the feasability of measuring the $B_s$ mixing parameter
$x_s$ = $\Delta m_s/\Gamma_s$. This measurement is key to obtaining the right
side of the unitarity triangle shown in Fig. 1. The value of $x_s$ may be
expected to fall in the range indicated in Fig. 4. This is a large range and
demostrates how hard it may be to measure. Recall that for $B_d$ mesons,
$x$ = 0.73. Thus the oscillation length for $B_s$ mixing is at least a factor
of 10 shorter and may approach a factor of 100!

\begin{figure}[htb]
\vspace{-.3cm}
\centerline{\epsfig{figure=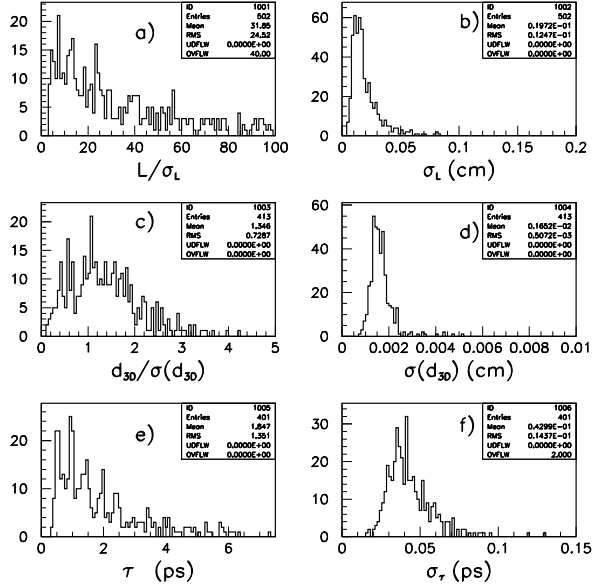,width=3.5in}}
\vspace{-.30cm}
\caption{\label{psiksanalcuts} Several analysis quantities determined for the
$\psi K^{*o}$ final state. $L/\sigma_L$ is the distance of the decay from
the primary divided by its error; $d_{30}$ refers to the distance of closest
approach of the $B$ direction to the primary (projecting backward along the
its momentum vector). $\tau$ refers to the lifetime of each event.}
\end{figure}
BTeV has investigated two final states which can be used. The first $\psi
K^{*o}$, $\psi\to \mu^+\mu^-$ and $K^{*o}\to K^-\pi^+$ has several advantages.
It can be selected using either a dilepton or detached vertex trigger.
Backgrounds can be reduced in the analysis by requiring consistency with the 
$\psi$ and $K^{*o}$ masses. Furthermore, it should have excellent time
resolution as there are four tracks coming directly from the $B$ decay vertex.
The one disadvantage is that the decays is Cabibbo suppressed, the Cabibbo
allowed channel being $\psi\phi$ which is useless for mixing studies. The
branching ratio therefore is predicted to have the low value of $8.5\times
10^{-5}$. 

Some important analysis quantities and their errors are shown in
Fig.~\ref{psiksanalcuts}. The lifetime resolution, $\sigma_{\tau}$ is
45 fs.

BTeV estimates that they will obtain 225 events/``Snowmass year". (A ``Snowmass
year has $10^7$ seconds.) The signal to background is 3.4:1. This probably
could be increased with little loss of efficiency by requiring that no other
tracks be consistent with the $B$ decay vertex, but this has not yet been done.

\begin{figure}[htb]
\vspace{-.03cm}
\centerline{\epsfig{figure=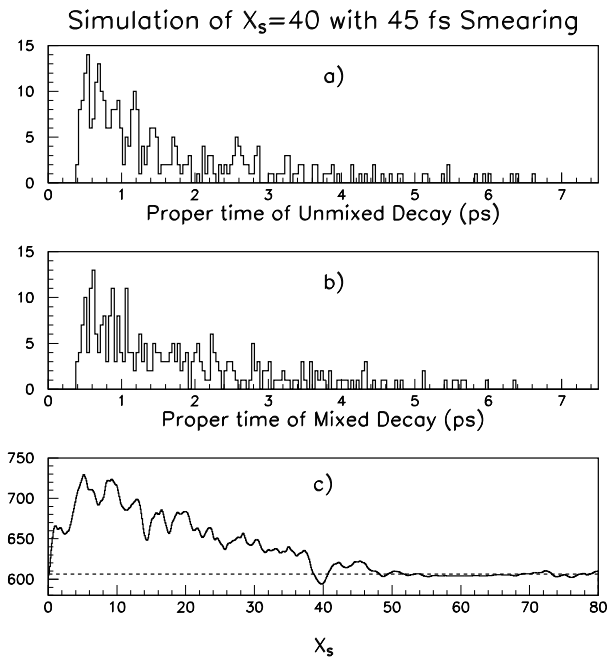,width=3.6in}}
\caption{\label{psikstime2} The observed decay time distributions for
$B_s\to\psi K^{*o}$ generated with $x_s = 40$. Unmixed decays are shown in
(a), mixed in (b). Background and miss tagging has been included. In (c) the
results of a likelihood fit to the time distributions are shown. The dashed
line shows a 5$\sigma$ difference from the best solution.}
\end{figure}

The time distributions of the unmixed and mixed decays are shown in
Fig.~\ref{psikstime2}, along with a calculation of the likelihood of there
being an oscillation as determined by fits to the time distributions.
Background and wrong tags are included. 
The fitting proceedure correctly finds the input value of $x_s=40$. The danger
is that a wrong solution will be found. The dashed line shows the change in
likelihood corresponding to 5 standard deviations. If our criteria is that the
next best solution be greater than $5\sigma$, then this is the best that can be
done with one years worth of data in this mode. Once a clean solution is found,
the error on $x_s$ is quite small, being $\pm 0.15$ in this case.

\begin{figure}[htb]
\vspace{-.03cm}
\centerline{\epsfig{figure=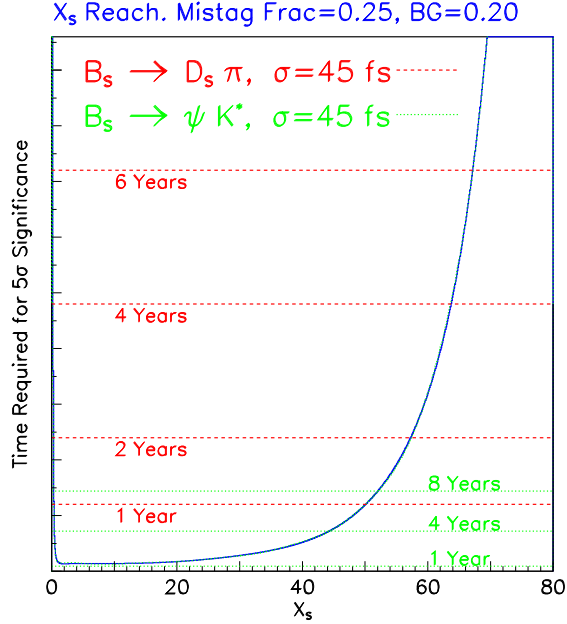,width=3.5in}}
\caption{\label{mix_summary} The $x_s$ reach for both $D_s^+\pi^-$ and
$\psi K^{*o}$ decays of the $\overline{B}_s$ The points at which the dashed
lines intersect the curve shows the number of years required to make a
significant measurement (see text) using $D_s\pi$, while the dotted lines
show the same thing for $\psi K^{*o}$.}
\end{figure}

BTeV has also investigated the $D_s^+\pi^-$ decay of the $\overline{B}_s$, with
$D_s^+\to\phi\pi^+$. It turns out that the lifetime resolution is 45 fs, the
same as for the $\psi K^{*o}$ decay mode. Since the predicted branching ratio
for this mode is 0.3\%, many more such events are obtainable in one year of
running. Fig.~\ref{mix_summary} shows the $x_s$ reach obtainable for a
$5\sigma$ discrimination between the favorite solution and the next best
solution, for both decay modes. The background is assumed to be 20\% and the
flavor mistag fraction is taken as 25\%. The tagging efficiency is taken as
10\%. The dashed lines show the number of years of running, where one year is
$10^7$ seconds.

The $x_s$ reach is excellent and extends over the entire predicted Standard
Model range. The curve gets very steep near $x_s$ values of 50 because the time
resolution is too poor to resolve the oscillations.

LHC-B has modeled 40 fs resolution in $D_s^+\pi^-$ and will be able to make
an excellent measurement of $B_s$ mixing.

\subsection{Measurement of the CP violating asymmetry in $B^o\to\pi^+\pi^-$}

\begin{figure}[htb]
\centerline{\epsfig{figure=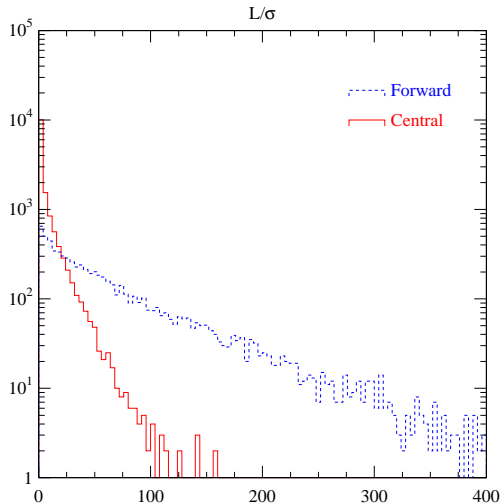,width=2.7in}}
\caption{\label{l_over_sig} Comparison of the $L/\sigma$ distributions
for the decay $B^o\to\pi^+\pi^-$ in central and forward detectors
produced at a hadron collider with a center of mass energy of 1.8~TeV.}
\vspace{1 cm}
\end{figure}

The trigger efficiency for this mode has already been discussed.
For the $B^o\to\pi^+\pi^-$ channel BTeV has compared the offline fully 
reconstructed decay length distributions in their forward geometry with that of
detector configured to work in the central region. 
Fig.~\ref{l_over_sig} shows the normalized 
decay length expressed in terms of $L/\sigma$ where $L$ is the decay length and 
$\sigma$ is the error on $L$ for the $B^o\to \pi^+\pi^-$ decay \cite{procario}.

 The forward 
detector clearly has a much more favorable $L/\sigma$ distribution, which is
due to the excellent proper time resolution. Being able to keep high efficiency
in the trigger and analysis levels and being able to decimate the backgrounds
relies mainly on having the excellent $L/\sigma$ distribution shown for the
forward detector.

\begin{figure}[htb]
\epsfig{figure=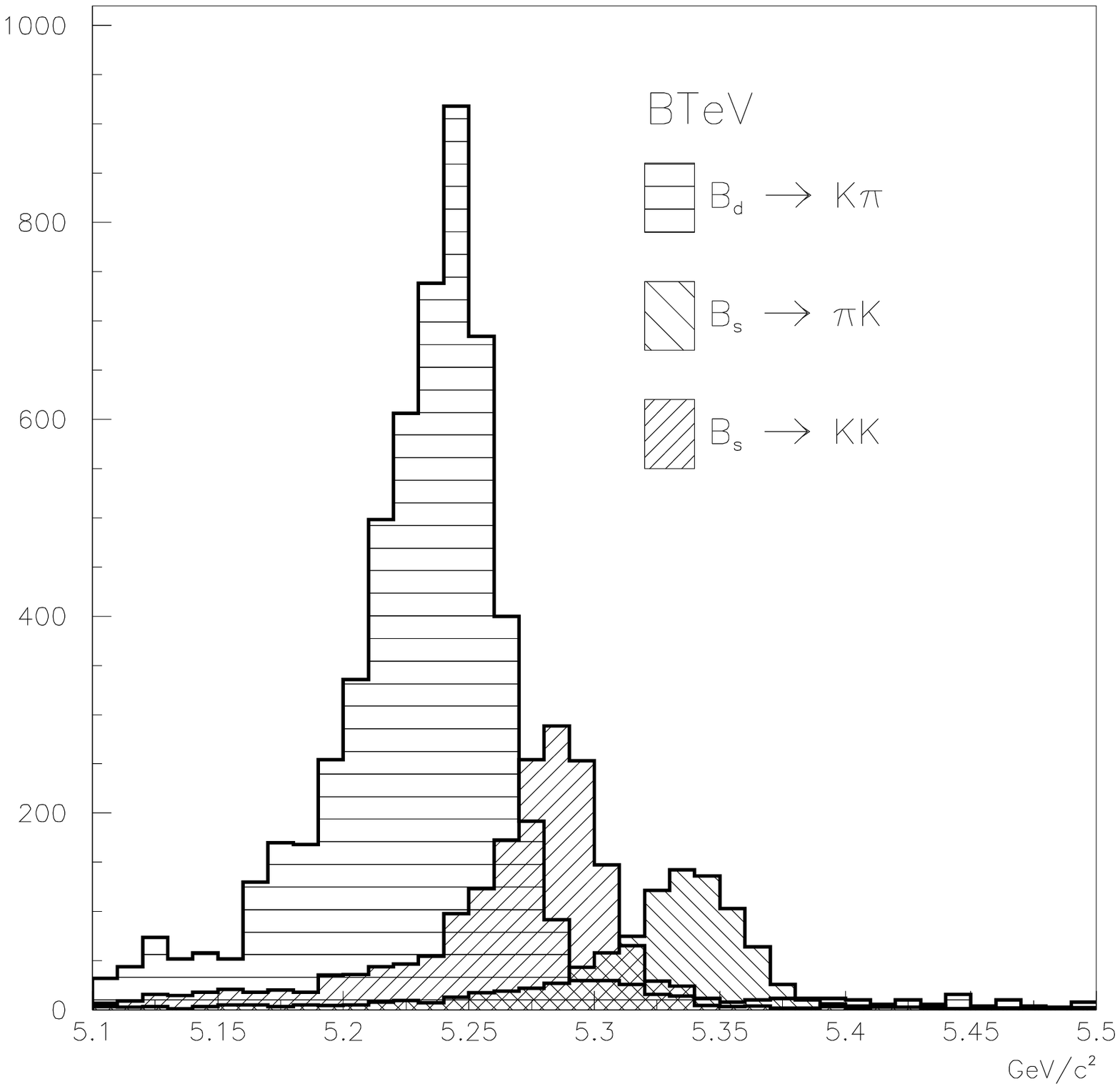,width=2.5in}\
\epsfig{figure=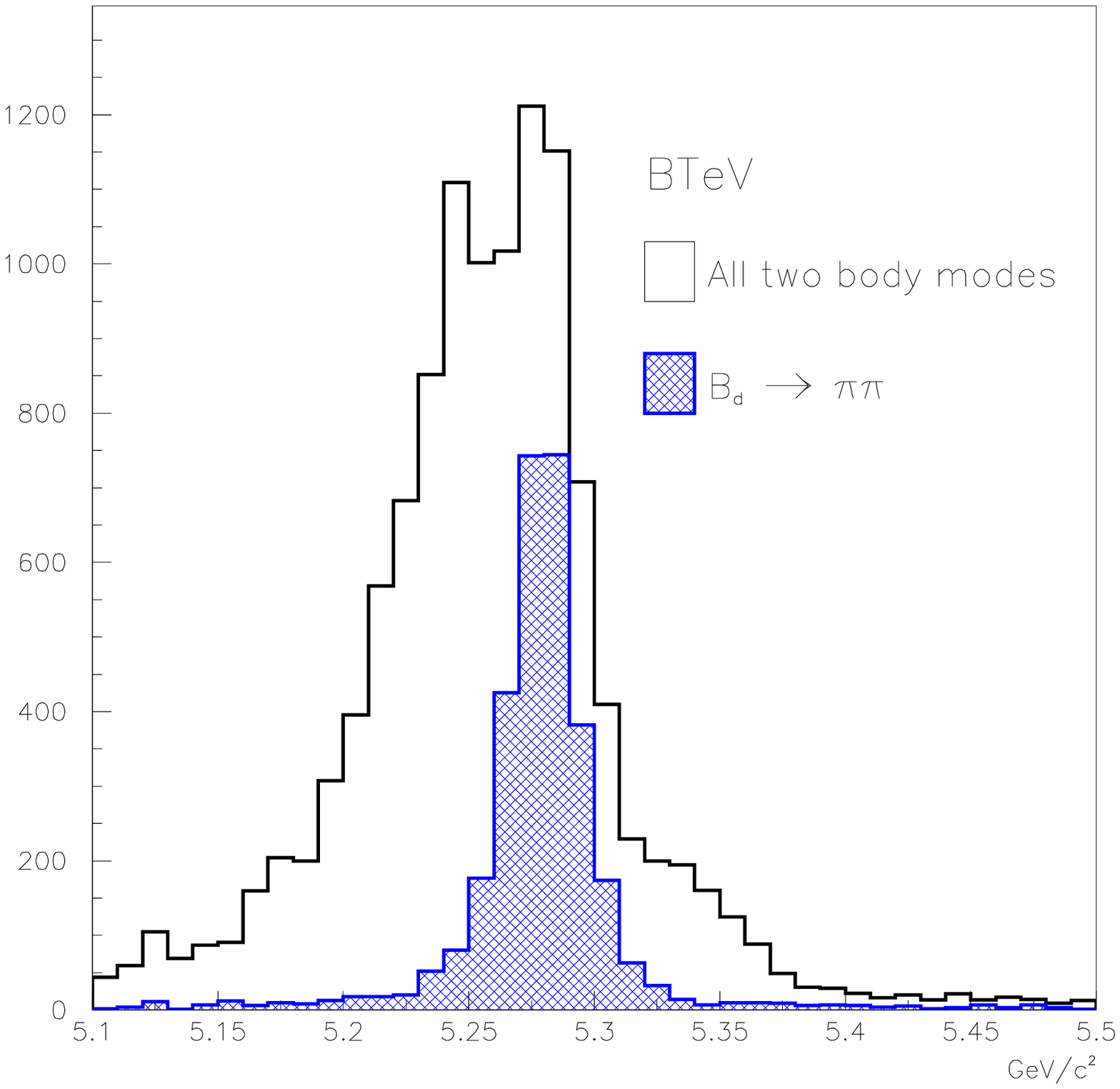,width=2.5in}
\caption{\label{pipi_nopid} Invariant mass distributions of all $B\to h^+h^-$
final states, where $h$ denotes either a pion or kaon, and the mass is computed
assuming that both tracks are pions. The plot on the left shows the individual
background channels and the one on the right shows the sum of all channels properly
normalized (see text) to the $\pi^+\pi^-$ signal.}
\vspace{1 cm}
\end{figure}

For this analysis $L/\sigma$ is required to be $>10$. Each pion track is
required to miss the primary vertex by a distance/error $>5\sigma$ and that the
$B^o$ point back to the primary by a distance/error $<2\sigma$. Furthermore,
each track is required to be identified as a pion and not a kaon in the RICH
detector. Without particle identification it is impossible to distinguish
$B^o\to\pi^+\pi^-$ from the combination of $B^o\to K^{\pm}\pi^{\mp}$, $B_s\to
K^+K^-$ and $B_s\to K^{\pm}\pi^{\mp}$, as is shown on Fig.~\ref{pipi_nopid}.
Here $\cal{B}$$(B^o\to K^{\pm}\pi^{\mp})$ is taken as $1.5\times 10^{-5}$ and
$\cal{B}$$(B^o\to \pi^+\pi^-)$ is taken as $0.75\times 10^{-5}$, from recent
CLEO measurements \cite{CLEOpp}. The $B_s$ decay into $K^+K^-$ is assumed to
have the same rate as the $B^o$ decay into $K^{\pm}\pi^{\mp}$, and the $B_s$
decay into $K^{\pm}\pi^{\mp}$ is assumed to have the same rate as the $B^o$
decay into $\pi^+\pi^-$.

Using the good particle identification, BTeV predicts that they can measure the
CP violating asymmetry in $\pi^+\pi^-$ to $\pm 0.05$ as detailed in
Table~\ref{tab:pipi}.

\begin{table}
\begin{center}
\begin{narrowtabular}{2cm}{lcc}\hline 
Quantity   &  Value    \\ \hline
Cross-section & 100 $\mu$b      \\
Luminosity & $0.5\times 10^{32}$ \\
\# of $B^o$/Snowmass year  & $3.8\times 10^8$ \\
$\cal{B}$$(B^o\to \pi^+\pi^-)$  & $0.75 \times 10{-5}$ \\
Reconstruction efficiency & 0.09 \\
Triggering efficiency (after all other cuts) & 0.72 \\
\# of $\pi^+\pi^-$ & 17000\\
$\epsilon D^2$ for flavor tags {\small($K^{\pm}$, $\ell^{\pm}$, same + opposite
sign jet tags)}   & 0.1 \\
\# of tagged $\pi^+\pi^-$ & 1700\\
Signal/Background & 0.4 \\
Error in asymmetry (including background) & $\pm 0.05$ \\ \hline 
\end{narrowtabular}
\caption{Numbers entering into the accuracy in measuring the CP violating
asymmetry in $B^o\to \pi^+\pi^-$. \label{tab:pipi}}
\end{center}
\end{table}

\subsection{Finding the rare decay $B^-\to K^-\mu^+\mu^-$}
Interesting rare decays include dimuon, dielectron and single photon channels.
The rates are largest among the single photon channels. However, the
feasability of trigger and analyzing these decays has not been attempted. The
branching ratios are larger for muons than electrons. They are given in
Table~\ref{tab:dimuon} \cite{dilepbranch}.

\begin{table}[t]
\caption{Rare dimuon decay branching ratios.\label{tab:dimuon}}
\vspace{0.4cm}
\begin{center}
\begin{tabular}{lccccc}   \hline
channel & $B^-\to K^-\mu^+\mu^-$ & $\overline{B}^o\to K^{*o}\mu^+\mu^-$ &
$B^-\to \mu^+\mu^-$ & $B_s\to \phi\mu^+\mu^-$ & $B_s\to \mu^+\mu^-$\\\hline
predicted $\cal{B}$ & $4\times 10^{-7}$ &$1.5\times 10^{-6}$ &
$1.5\times 10^{-10}$ &$1.5\times 10^{-6}$ & $3.5\times 10^{-9}$ 
 \\ \hline
\end{tabular}
\end{center}
\end{table}

The final state we are discussing has a significantly lower branching ratio
than available in $B_s\to \phi\mu^+\mu^-$. Furthermore, the narrow $\phi$ is
useful for background suppression. Thus this mode should be easier to detect.

While tagging is not needed for this physics, the backgrounds are a problem. To
detect these rare final states, a veto must be put on final states which are
decays of the $\psi$ or $\psi'$, which are much more prolific. Similar cuts are
applied as used when studying $B^o\to\pi^+\pi^-$. It is interesting to note
that requiring kaon identification in the RICH reduces the background by a
factor of three without significantly reducing the efficiency. These events can
be triggered on either the detached vertex or the dimuon lines. The analyis
efficiency is 2.5\% and the trigger efficiency on events which pass the
analysis cuts is 85\%. BTeV finds 300 signal events using an integrated
luminosity of 500 pb$^{-1}$ with a signal to background of 0.4.

\section{Comparisons}
\subsection{Comparison with $e^+e^-$ $B$ factories}
Let us compare one example of the physics reach of a foward collider detector
(BTeV)  with $e^+e^-$ $B$ factories operating at the $\Upsilon$(4S).
Table~\ref{tab:pipic} shows the effective number of flavor tagged
$B^o\to\pi^+\pi^-$ events found in a Snowmass year of running. The branching
ratio is assumed to be $0.75\times 10^{-5}$.
\begin{table}[t]
\caption{Number of tagged $B^o\to\pi^+\pi^-$ \label{tab:pipic}}
\vspace{0.4cm}
\begin{center}
\begin{tabular}{lccccrr}\hline
 & $\cal{L}$(cm$^{-2}$s$^{-1}$) & $\sigma$ & \#$B^o/10^7$s  & efficiency &
 $\epsilon D^2$ & \# tagged
\\\hline
$e^+e^-$ & $3\times 10^{33}$ & 1 nb & $3.0\times 10^7$ & 0.4 & 0.4 & 46 \\
BTeV & $5\times 10^{31}$ & 100 $\mu$b & $3.5\times 10^{10}$ & 0.065 & 0.1 & 1700
 \\ \hline
\end{tabular}
\end{center}
\end{table}
Clearly the hadron collider experiment has the ability to collect far more
events.

\subsection{Comparison of BTeV with Tevatron central detector}
There are several important areas where BTeV has crucial advantages over a
Tevatron central detector.
\begin{itemize}
\item {\it The detached vertex trigger.-} The ability to trigger on decay
verticies allows BTeV to address important issues in charm physics including
mixing and CP violation. It allows accumulation of a plethora of interesting
$B$ decay states including semileptonic decays and hadronic decays such as
$B^o\to \pi^+\pi^-$ and $\overline{B}_s^o\to D_s^+ h^-.$
\item {\it Resolution on detached vertex.-} The $L/\sigma_L$ distribution as
shown in Fig.~\ref{l_over_sig} is much more favorable for the forward
detectors. This allows for high efficiency for detecting the final state of
interest and permits good background rejection.
\item {\it Charged hadron identification.-} Crucial for measurement of many
final states and CP asymmetries. For example, determining the CP asymmetries
in $B^o\to\pi^+\pi^-$, $B_s\to K^+K^-$, $B_s\to D_s^{\pm} K^{\mp}$. It also
provides a larege increase in flavor tagging efficiency which helps to
precisely measure all asymmetries.
\end{itemize}

\subsection{Comparison of LHC-B with BTeV}
There are issues favoring both LHC-B and BTeV:
\begin{itemize}
\item {\it Issues favoring LHC-B}
\begin{itemize}
\item The $b$ cross-section is expected to be five times larger at LHC than at
the Tevatron.
\item The mean number of interactions per beam crossing is four times lower at
LHC than at the Tevatron even when the bunch spacing at FNAL is 132 ns.
\end{itemize}
\item {\it Issues favoring BTeV}
\begin{itemize}
\item BTeV is a two-arm spectrometer which increases the physics yield by a
factor of two relative to LHC-B.
\item The 25 ns bunch spacing at LHC makes first level detached vertex 
triggering more difficult than at the Tevatron with 132 ns bunch spacing.
\item The seven times larger LHC beam energy causes problems. There is a much
larger range of track momentua that need to be momentum analyzed and
identified, and there is a large increase in track multiplicity which causes
difficulties in triggering and pattern recognition.
\item BTeV is designed with the vertex detector in the magnetic field in order
to reject low momentum tracks at the trigger level. Low momentum tracks can
multiple scatter and cause false triggers. 
\item The detached vertex trigger in BTeV allows for charm physics studies.
\end{itemize}
\end{itemize}

\section{Conclusions}
The BTeV and LHC-B programs, emphasizing studies of mixing, CP violation and
rare decays offer exciting and unique physics opportunities prior to (BTeV)
and in the LHC era (BTeV and LHC-B). Standard Model parameters will be
precisely measured and physics beyond the Standard Model may appear.

\section*{Acknowledgments}
I would like to thank my BTeV colleagues including Marina Artuso, Joel Butler,
Chuck Brown, Paul Lebrun, Patty McBride, Mike Procario and Tomasz Skwarnicki 
for their help in getting this material together. This work was supported by
the U. S. National Science Foundation.

\end{document}